\documentclass[aps,prl,reprint,superscriptaddress]{revtex4-2}
\usepackage{graphicx}
\usepackage{amsmath}
\usepackage{braket}
\usepackage{bm}

\begin{document}


\title{Robust spin-transfer torque and magnetoresistance in non-collinear antiferromagnetic junctions}


\author{Srikrishna Ghosh}
\affiliation{Institute of Physics, Czech Academy of Sciences, Cukrovarnick\'{a} 10, 162 00 Praha 6 Czech Republic}
\author{Aurelien Manchon}
\affiliation{CINaM, Aix-Marseille Univ, CNRS, Marseille, France}
\author{Jakub \v{Z}elezn\'{y}}
\affiliation{Institute of Physics, Czech Academy of Sciences, Cukrovarnick\'{a} 10, 162 00 Praha 6 Czech Republic}


\date{\today}

\begin{abstract}
 Ferromagnetic spin-valves and tunneling junctions are crucial for spintronics applications and are one of the most fundamental spintronics devices. Motivated by the potential unique advantages of antiferromagnets for spintronics, we theoretically study here junctions built out of non-collinear antiferromagnets. We demonstrate a large and robust magnetoresistance and spin-transfer torque capable of ultrafast switching between parallel and anti-parallel states of the junction. In addition, we show that the non-collinear order results in a spin-transfer torque that is in several key aspects different from the spin-transfer torque in ferromagnetic junctions.
\end{abstract}
 

\maketitle

In ferromagnetic (FM) materials the spins of electrons that carry electrical current are preferentially oriented along one direction: the electrical current is spin-polarized. When this spin-polarized current is injected into a different FM with a misaligned magnetization orientation, the spin-polarization has to reorient. Due to angular momentum conservation, this exerts a torque on the FM, known as the spin-transfer torque (STT) \cite{Brataas2012,Ralph2008}. This torque can be used to switch magnetization in a FM or to move magnetic domain walls. It is typically utilized in nanoscopic devices composed of two thin ferromagnetic layers separated by a metallic or insulating spacer. These devices are known as spin-valves in the metallic case or magnetic tunnel junctions in the insulating case and will be referred to as junctions in the following. STT can be used to electrically switch between a parallel and anti-parallel configuration of the junctions. The two configurations can in turn be distinguished by the giant magnetoresistance (GMR) \cite{Baibich1988,Binasch1989} or the tunnel magnetoresistance effects (TMR) \cite{JULLIERE1975225}. The GMR and TMR are widely used as magnetic sensors in hard disk drives and the STT together with the GMR or the TMR form the basis of the novel type of magnetic memories, the magnetic random access memories (MRAMs) \cite{Apalkov2016,Khvalkovskiy2013}. 

In this letter, we consider metallic junctions composed of AFMs instead of FMs. AFMs can offer significant advantages over FMs for spintronics applications \cite{Jungwirth2015,Baltz2018,Jungwirth2018,Zelezny2018}. Their magnetic dynamics is several orders of magnitude faster, allowing for much faster switching. The lack of a net magnetic moment implies no stray field, thus possibly allowing for closer packing of individual bits. Furthermore, a large variety of antiferromagnetic materials exists, such as insulators and semiconductors, multiferroics \cite{Eerenstein2006} or superconductors \cite{Lu2015}. FM materials accommodating such exotic electronic properties are much less common in nature. A number of theoretical works have shown that the STT, as well as the GMR or TMR, can also exist in an AFM junctions \cite{Nunez2006,Haney2007,Merodio2014,Stamenova2017,Zelezny2018}.  In contrast to the FM case, however, such effects arise from quantum-coherent scattering \cite{Nunez2006} and as a result, are very sensitive to the presence of disorder \cite{Duine2007,Saidaoui2014,Manchon2017,Zelezny2018}. This has a fundamental reason: in the simple AFMs that were primarily studied, the electrical current is not spin-polarized and consequently the STT has to vanish in the semiclassical limit \cite{Manchon2017}. It was found, however, that in the insulating case the STT can be more robust \cite{Saidaoui2016a}, although the reasons for this are not fully understood. Experimentally, no clear evidence of a STT, GMR or TMR in an AFM junction has been found so far \cite{Zelezny2018}. Instead, AFM spintronics has mainly focused on relativistic effects, which allow for both torque and magnetoresistance but are often smaller than the non-relativistic effects such as the STT \cite{Zelezny2018}.

Recently, it was discovered that in some types of AFMs spin-polarized current can exist. This was first found for non-collinear AFMs (AFMs in which the individual moments are not oriented along a single axis) such as Mn$_3$Ir or Mn$_3$Sn \cite{Zelezny2017b}  but later also found to exist in collinear AFMs  such as RuO$_2$ or MnF$_2$ \cite{Naka2019,Hernandez2021,Ahn2019,Hayami2019,Smejkal2020,Yuan2020}. Such spin-polarized currents are directly analogous to the spin-polarized currents in FMs. It suggests that a robust STT, as well as the GMR or TMR effects, could exist in junctions composed of AFMs in which the spin-polarized current exists \cite{Zelezny2017b,Smejkal2021giant}. 

Motivated by these results we theoretically study in this work the STT and GMR in non-collinear AFM junctions. Apart from the AFM order, these junctions differ from the well-known FM ones in their non-collinear nature. As we will show this plays an important role for the STT. The main goal of our work is to understand the general properties of STT and GMR in non-collinear AFM systems and to explore the role of disorder. We consider a relatively simple 2D model, which is not meant to quantitatively describe realistic systems but which covers the most important features of non-collinear coplanar antiferromagnets of interest to experiments (for example Mn$_3$Ir or Mn$_3$Sn). Importantly, this model allows for a comprehensive study of the role of disorder. In addition to point defects, we include a disorder representing structural imperfections that invariably exist in real systems but have not been included in previous studies and are expected to be important in simple antiferromagnets \cite{Zelezny2018}. For comparisons, we calculate also the GMR and STT in FM junctions.

Our simulations show that the STT in the non-collinear AFM junctions originates from the absorption of the spin-polarized current, analogously to the FM junctions. Consequently, we find that the STT, as well as the GMR, have a similar magnitude and robustness against disorder in the non-collinear AFM junctions as in the FM junctions. Using magnetic dynamics simulations we also demonstrate that the STT can switch the non-collinear AFM junction between parallel and anti-parallel states. Thus the AFM junctions have the same basic functionality as the FM junctions.

In addition, we find that the STT in non-collinear AFMs differs from collinear FMs in several key and general aspects stemming from the non-conservation of spin in non-collinear systems. One consequence is that  in the non-collinear junctions the torque is present for any relative orientation of the two magnetic layers. In contrast, in  FM junctions the torque vanishes for the parallel or anti-parallel configurations, which implies that thermal fluctuations are required to assist the magnetization switching, resulting in stochasticity and slower reversal time. We also find that a locally generated torque appears in each magnetic layer that cannot be attributed to the spin-polarized current from the other magnetic layer and is instead associated with a locally generated spin-polarized current. We find that this torque is typically large and that it has an angular dependence that is distinct from the well-known anti-damping torque in FMs.


\begin{figure}
\includegraphics[width=\columnwidth]{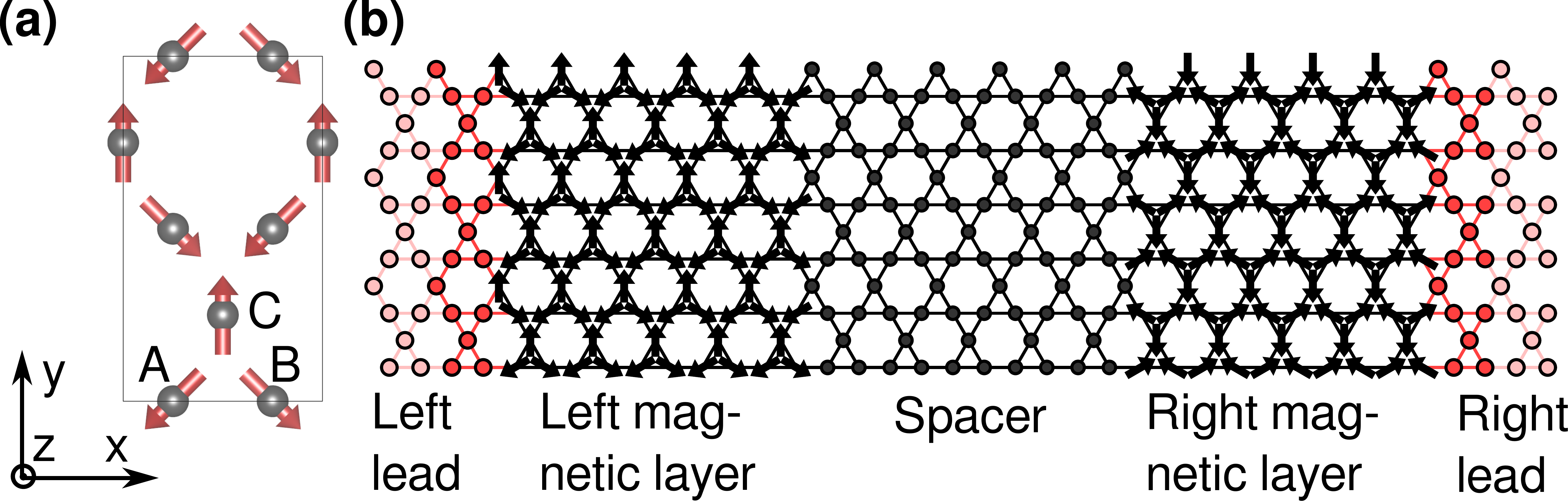}%
\caption{\label{fig:structure} \textbf{(a)} The unit cell of the model.  \textbf{(b)} Illustration of the junction. The left and right leads are periodic and semi-infinite along the x directions. The junction is finite along the $y$ direction. In the calculation the width along $y$ is much wider than illustrated here.}
\end{figure}

 We consider a 2D AFM system, shown in Fig. \ref{fig:structure}. The system is hexagonal with three atoms in a unit cell; however, for easy construction of the junctions, we double the unit cell, which makes the unit cell rectangular. Similar magnetic order exists in real materials, such as the Mn$_3$X AFMs \cite{Tomiyoshi1982,Yamaoka1974,Zelezny2017b}. The model we use consists of conduction $s$-electrons coupled to on-site magnetic moments. This model has been utilized in previous studies of non-collinear AFMs \cite{Chen2014,Zhang2018,Hayami2020}. We do not include spin-orbit coupling in our model, since the aim is to explore non-relativistic effects, in analogy to the FM junctions. In absence of disorder the system is described by the following Hamiltonian:
 \begin{align}
  H = t \sum_{<ab>\alpha} c^\dagger_{a\alpha}c_{b\alpha} + J \sum_{a\alpha,\beta} ({\boldsymbol \sigma} \cdot \mathbf{m}_a)_{\alpha\beta}c^\dagger_{a\alpha}c_{a\beta}.
 \end{align}
 Here, $c^\dagger$ and $c$ denotes the creation and annihilation operators respectively, $a,b$ denote the site index, $\alpha,\beta$ the spin index. The first term is the nearest neighbor hopping term, with $t$ representing the hopping magnitude. The second term represents the coupling of the conduction electrons to the on-site magnetic moments. Here $\mathbf{m}_i$ is the magnetic moment direction, $J$ is the exchange parameter and ${\boldsymbol \sigma}$ is  the vector of Pauli matrices. We always set $t=1\ \text{eV}$ and $J=-1.7\ \text{eV}$.
 
 To describe disorder we include for each site on-site energy chosen randomly from a gaussian distribution centered at 0 and with a standard deviation $D$. We describe the quantum transport using the scattering (Landauer-B\"uttiker) formalism as implemented in the Kwant package \cite{Groth2014}. We consider a system composed of two magnetic layers separated by a non-magnetic spacer (scattering region) and attach to the system perfectly periodic semi-infinite leads as illustrated in Fig. \ref{fig:structure}(b). The transport properties of the system are described by scattering of the incoming states from the left lead to the outgoing states in the left and right leads. 
 
 We use both magnetic and non-magnetic leads in the calculations. In the magnetic case, we use the same magnetic order for the left and right leads as for the left and right magnetic layers respectively. In such a case the calculation describes a junction with thick magnetic layers, whereas in the case of non-magnetic leads the calculations describe systems with thin magnetic layers. For simplicity, we discuss only the magnetic leads case in the main text, but we give the results for the non-magnetic leads in the Supplemental Material. In general, we find that for the FM junctions the results for non-magnetic leads are very similar, whereas for the non-collinear AFM there can be significant differences. The qualitative behavior is still generally the same, however. We note that the disorder is included only in the scattering region. We always set the width of the magnetic layers and the spacer along the $x$ direction to 5 unit cells each. The width of the system along the $y$ direction is set to 50 unit cells, which ensures that the effect of the top and bottom interfaces is negligible. To describe interface roughness we randomly include atomic interfacial steps using a random walk along the interfaces between the magnetic leads and the spacer as described in the Supplemental Material. The magnitude of the interfacial disorder is controlled by the parameter $n_\text{steps}$, which determines the average number of the interfacial steps.

 Within the scattering formalism, the transport is due to the difference of the chemical potentials of the left and the right leads $\delta \mu$. For the case of the electric effects, this difference is due to applied voltage $V$: $\delta \mu = -e V$. We always assume transport from the left to the right and also assume that the voltage is small, i.e. we will assume that the response to the electric field is linear. The current induced by the voltage is given by $I = GV$, where
 \begin{align}
 G = \frac{e^2}{h} \sum_{nm} |t_{nm}|^2.
 \end{align}
 Here, $t_{nm}$ is the transmission amplitude from incoming state $n$ in the left lead to outgoing state $m$ in the right lead. Using the scattering wavefunctions $\psi_n$ inside the lead associated with the $n$-the incoming states of the left lead at energy $E_F$ we can evaluate response of any quantity.  For an observable $A$ represented by operator $\hat{A}$ we have $\delta A = \chi_A \delta \mu$, where
 \begin{align}
  \chi_A = \sum_n \bra{\psi_n}\hat{A}\ket{\psi_n}. 
 \end{align}
 Note that we assume here a zero temperature Fermi-Dirac distribution. We are primarily interested in the torque. The torque acting on a site $a$ can be calculated using the torque operator $\hat{\mathbf{T}}_a = -J\sum_{\alpha\beta}(\mathbf{m_a} \times  {\boldsymbol \sigma})_{\alpha\beta}c^\dagger_{a\alpha}c_{a\beta}$. As we discuss in the Supplemental Material, the torque can also be evaluated from the non-equilibrium spin accumulation and, since we do not consider spin-orbit coupling the torque is also directly related to spin current: the torque on a site is given by the spin source at this site. We note that although the spin of the conduction electrons is strongly non-conserved in the non-collinear systems, spin current is nevertheless well defined in the non-relativistic limit since there is no conversion of the spin angular momentum to orbital angular momentum. 
 
\begin{figure}
    \includegraphics[width=\columnwidth]{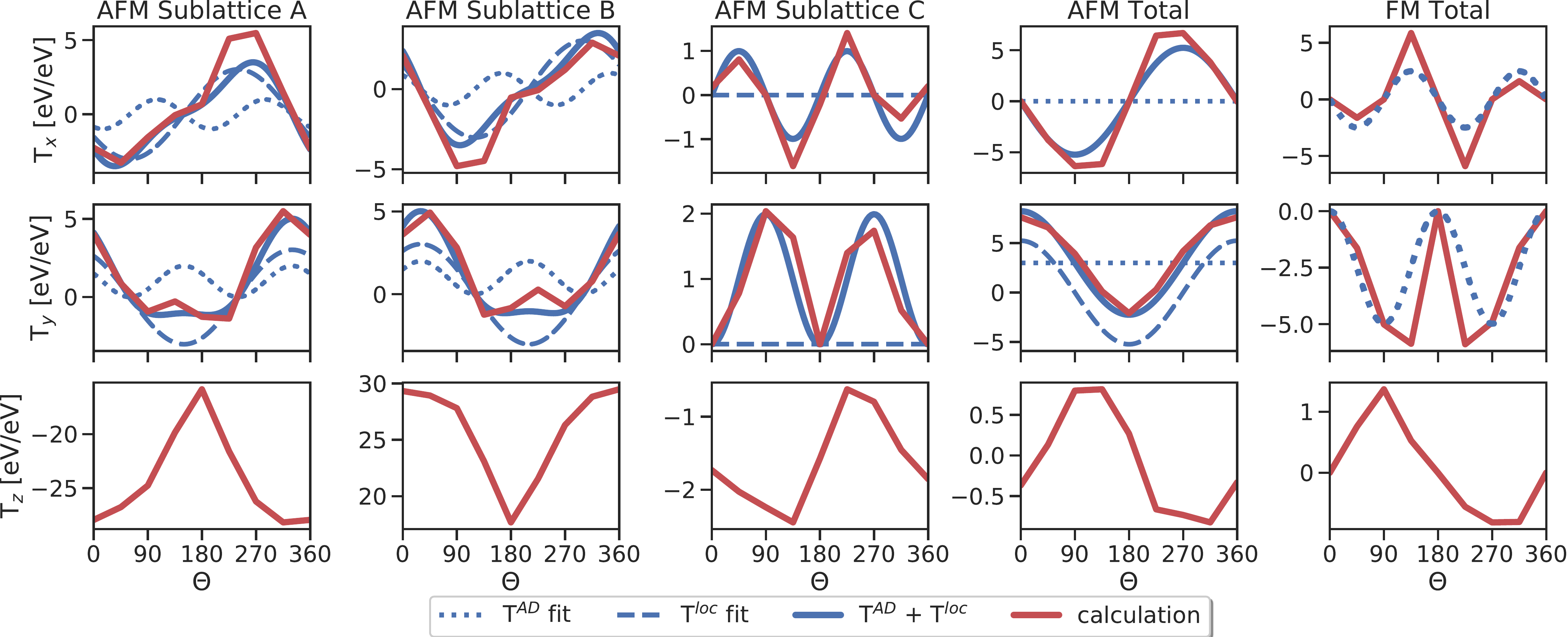}%
    \caption{\label{fig:sublattice_angular_dependence}  The dependence of sublattice and the total torque in the right magnetic layer as a function of the rotation angle of the right magnetic layer for the AFM and FM junctions. The in-plane components of the torque are fitted by the combination of $\mathbf{T}^\text{loc}$ and $\mathbf{T}^\text{AD}$. Here $E_F = -1.25\ \text{eV}$, $D=0.2\ \text{eV}$ and $n_\text{steps} = 25$. The torque has units of eV as is given per the applied bias, which is also given in units of eV.}
\end{figure}

In Fig. \ref{fig:sublattice_angular_dependence} we give the calculated torque in the right magnetic layer for FM and AFM junctions as a function of angle $\theta$, which denotes the rotation of the right magnetic layer in the $x-y$ plane, starting from a parallel junction. Here we set $D=0.2\ \text{eV}$ and $n_\text{steps} = 25$. We find that in the FM case, the main term of the torque is the well-known anti-damping torque $\mathbf{T}^\text{AD} \sim \mathbf{m} \times (\mathbf{m} \times \mathbf{p})$, where $p$ denotes the spin-polarization of the spin current, which in our case is determined by the left magnetic layer and is oriented along the $y$ direction. It has been considered previously that in AFMs, the spin-transfer torque has the anti-damping form on each sublattice \cite{Gomonay2012,Cheng2014c}: $\mathbf{T}_a^\text{AD} \sim \mathbf{m}_a \times (\mathbf{m}_a \times \mathbf{p})$, where $a$ denotes the sublattice  (dotted blue line in Fig. \ref{fig:sublattice_angular_dependence} ). We find, however, that in our case the torque cannot be understood only in this way. As a consequence of the non-collinearity and the broken inversion symmetry by the interface, a locally generated torque appears in the right magnetic layer that cannot be attributed to spin-transfer from the left magnetic layer (and vice versa). Such torque is associated with a spin-polarized current generated in the right magnetic layer. This spin-polarized current is in our case always polarized along the direction of the magnetic moments on sublattice $C$, thus we can expect in analogy with the anti-damping torque: $\mathbf{T}^\text{loc}_a \sim \mathbf{m}_a \times (\mathbf{m}_a \times \mathbf{m}_C)$  (dashed blue lines in Fig. \ref{fig:sublattice_angular_dependence}). As shown in Fig.  \ref{fig:sublattice_angular_dependence}, the in-plane components of the calculated torque in the AFM can indeed be modelled reasonably well by the combination of $\mathbf{T}^\text{AD}$ and $\mathbf{T}^\text{loc}$  (solid blue lines in Fig. \ref{fig:sublattice_angular_dependence}). As discussed in the Supplemental Information, we find that this is typically the case also for other parameters.

In addition to the in-plane torque, which is well described by the combination of $\mathbf{T}^\text{AD}$ and $\mathbf{T}^\text{loc}$, we also find a $T_z$ component of the torque. In FMs this torque typically has a field-like character, which we also find here, in both the FM and the total torque in the AFM. This torque occurs because the spin current that is reflected from the right magnetic layer contains a $z$ polarized component when the two magnetic layers are misaligned.

\begin{figure}
    \includegraphics[width=\columnwidth]{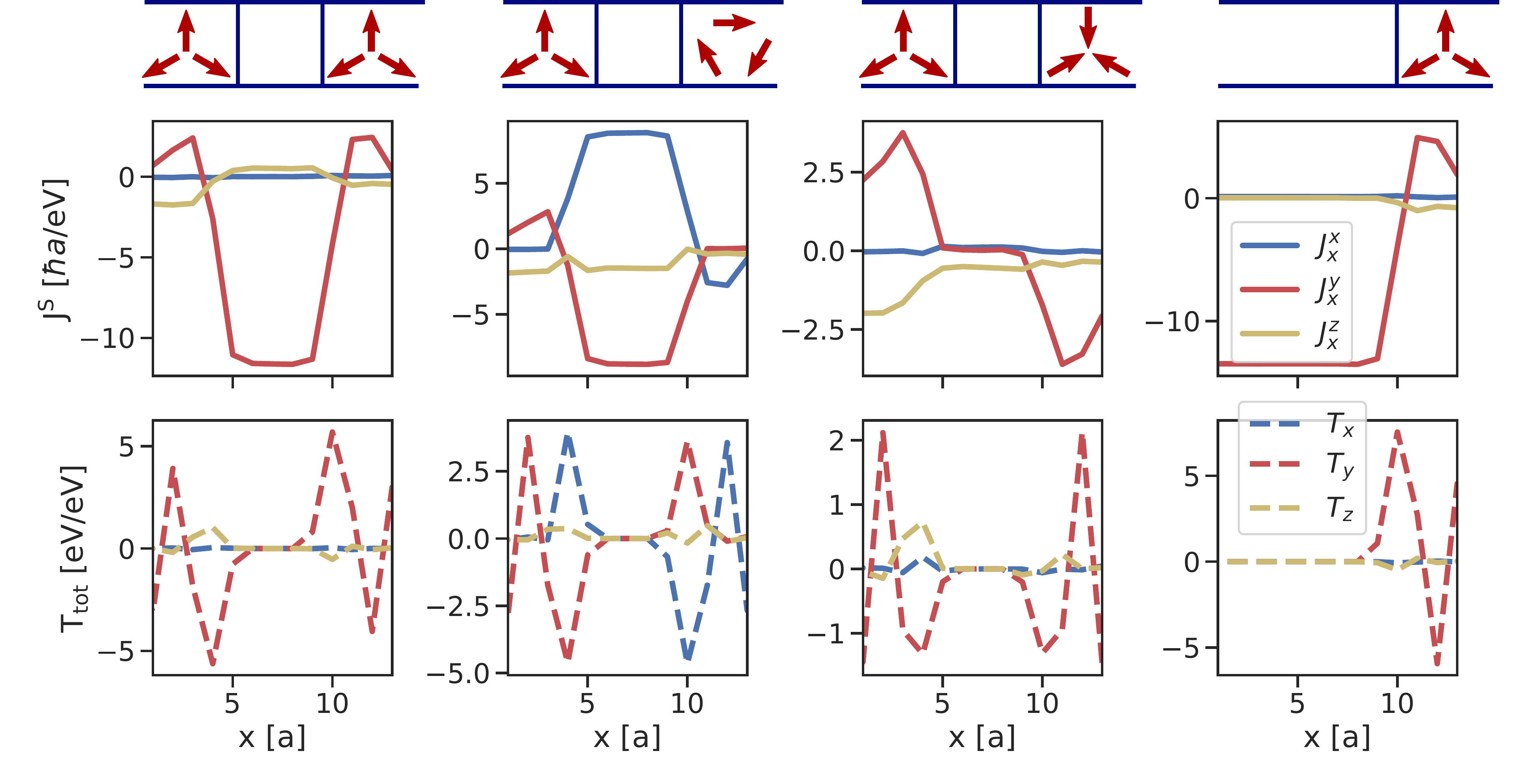}%
    \caption{\label{fig:x_profile} The spin current and torque for the parallel, perpendicular, and anti-parallel configurations of the AFM junction as well for a junction composed of only a non-magnetic and one AFM layer as a function of the $x$ coordinate ($a$ denotes the lattice constant along the $x$ direction). We sum up both the torque and the spin current within each unit cell and then sum up also along the $y$ direction. $J_x^i$ denote spin current flowing along the $x$ direction with spin-polarization along $i$. We set $D = 0.2\ \text{eV}$, $n_\text{steps}=25$ and $E_F=-1.25\ \text{eV}$.}
\end{figure}

The connection between the spin-polarized current and the torque is illustrated in Fig. \ref{fig:x_profile}, where we show the spin current and the total torque within the junction as a function of the $x$ coordinate. We see that for all the configurations of the junction the torque is directly connected to absorption or generation of the spin-polarized current. In the FM case no torque or, equivalently, no spin source can be present in the parallel or anti-parallel configurations of the junction or in a junction containing only one FM layer since in such a case spin is conserved. In contrast, in the non-collinear case, we find a non-vanishing torque in any configuration because in the non-collinear AFM spin is never conserved. In the parallel and anti-parallel cases both $\mathbf{T}^\text{loc}$ and $\mathbf{T}^\text{AD}$ contribute since in the non-collinear system $\mathbf{T}^\text{AD}$ is non-zero for some sublattices for any orientation of the junction and will always sum up to non-zero total torque. We note that in addition to the torque due to a global spin current illustrated in Fig. \ref{fig:x_profile}, also a torque due to local spin currents can occur. This is the case of the large $T_z$ torque for the $A$ and $B$ sublattices illustrated in Fig. \ref{fig:sublattice_angular_dependence}.

The origin of the $\mathbf{T}^\text{loc}$ torque is illustrated by the case of the junction with only one magnetic layer, shown in Fig. \ref{fig:x_profile} since here the $\mathbf{T}^\text{AD}$ is absent. We see that in this case the single magnetic layer acts as a spin source and thus generates a spin-polarized current and a torque. This self torque is local in the sense that it is generated directly in the right magnetic layer.  However, since it is associated with a spin-polarized current, in a junction with two magnetic layers it can also influence the other layer. In other words, the $\mathbf{T}^\text{AD}$ torque in the right magnetic layer is due to a spin-polarized current from the left magnetic layer, which is associated with the  $\mathbf{T}^\text{loc}$ torque in the left magnetic layer (and vice versa). We note that in the FM case, a single FM layer also generates a spin-polarized current, however, this spin current is polarized along the FM magnetization direction and thus does not generate any torque. This also means that in such a case no spin source is present and therefore the spin current throughout the system is everywhere homogeneous because the reflected spins precisely balance the transmitted spins.

We use the calculated torque to simulate switching of the AFM junction. As discussed in the Supplemental Material, we find that both the in-plane torque characterized by a combination of  $\mathbf{T}^\text{loc}$ and $\mathbf{T}^\text{AD}$, and the $T_z$ torque can deterministically switch the junction between parallel and anti-parallel states using current pulses with opposite directions. Crucially, the switching is ultrafast (on a ps timescale) since in AFMs the dynamics is enhanced by the exchange interaction. This comes into play since the torque initially slightly cants the magnetic moments, which results in a large exchange torque. Unlike in the FM case, where thermal activation is necessary to activate the switching since no torque exists in the parallel state or anti-parallel states, we find that in the AFM case switching is possible directly from the parallel or anti-parallel states since the torque is always present. We note that $\mathbf{T}^\text{loc}$, which is typically the dominant term in our calculations cannot by itself be used for deterministic switching. This is because this torque is non-relativistic and internally generated which means that when the magnetic order is rotated the torque is rotated in the same way and can thus never vanish. As a result, the other magnetic layer is crucial here. Our simulations show that when $\mathbf{T}^\text{loc}$ is combined with $\mathbf{T}^\text{AD}$, deterministic switching is possible and that the $\mathbf{T}^\text{loc}$ can reduce the $\mathbf{T}^\text{AD}$ necessary for switching. The $\mathbf{T}^\text{loc}$ torque could also be used for spin-torque oscillators.

\begin{figure}
    \includegraphics[width=\columnwidth]{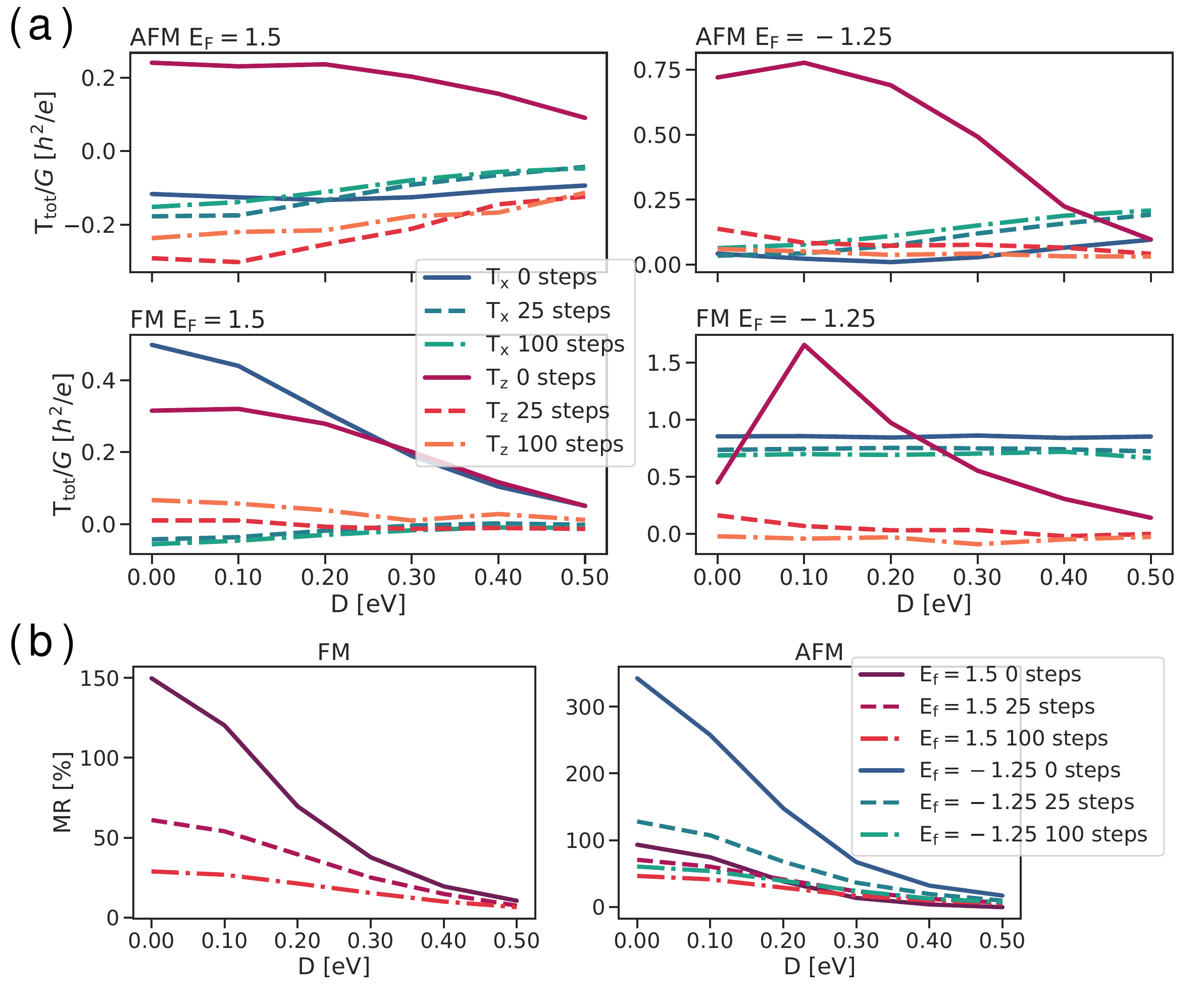}%
    \caption{\label{fig:torque_magnitude} \textbf{(a)} The dependence of the torque magnitude scaled by the conductance on the disorder paramater $D$ and on the interfacial disorder characterized by number of interfacial steps for the FM and AFM junction and for two values of $E_F$. The $T_y$ component is not shown here since it has a similar dependence on magnitude as $T_x$. For $T_z$ we set $\theta=90^\circ$; for $T_x$ $\theta=135^\circ$ for the  FM and $\theta=90^\circ$ for the AFM. \textbf{(b)} The dependence of GMR on disorder for the FM and AFM junctions. }
\end{figure}

In Fig. \ref{fig:torque_magnitude}(a) we show the dependence of the torque magnitude on the on-site and interfacial disorder for the FM and AFM junctions. For simplicity we give here the dependence for the total torque; however, the conclusions are generally the same when the sublattice torque is considered. We give the results here for two values of the Fermi level and we also scale the torque by the conductance since the torque magnitude per current density is the main quantity for practical utilization of the torque. Overall, we find that the torque in the AFM junctions has a similar magnitude and robustness against disorder as in the FM junctions. In agreement with previous considerations, we find that the interfacial steps reduce strongly the $T_z$ component of the torque, however, the $T_x$ and $T_y$ components are more robust and survive even with significant interfacial disorder present. We also find no strong reduction of the torque magnitude with the disorder parameter $D$. We note that the case of FM with $E_F = 1.5~\text{eV}$ is somewhat an exception as we find that in this case the torque is more sensitive to disorder than in the other cases. As shown in the Supplemental Material, the angular dependence of the torque is also generally unchanged with the disorder.



In Fig. \ref{fig:torque_magnitude}(b) we show the dependence of the GMR on disorder for the FM and AFM junctions. We find that in both cases the GMR is strongly reduced by disorder, but non-negligible GMR is present even in the presence of significant disorder. Crucially the robustness of the GMR is similar in the AFM as in the FM. An important difference is, however, that in the collinear FM it can happen that the  conduction electrons all have the same direction of spin, a so-called half-metallic system. In such a case the GMR is in the non-relativistic and zero-temperature limit infinite since no current can then flow in the anti-parallel state. This is what happens for our model for the case of $E_F = -1.25~\text{eV}$. In a non-collinear AFM, this is likely not possible since in such a system spin is never conserved. This could be a disadvantage of non-collinear systems. However, in practice spin is never fully conserved and most FMs that are used for GMR or TMR are not half-metals; thus we expect that the non-collinear AFMs can in practice have comparable GMR to the commonly used FMs.


We conclude the manuscript by discussing the generality of our results. Since our calculations confirm  that a strong link between the STT and the spin-polarized current exists, we expect that most AFMs in which current is spin-polarized  including the collinear ones will exhibit similarly robust STT. On the other hand, the features that distinguish the STT in the non-collinear AFMs from FMs, such as the local torque, originate from the strong non-conservation of spin in non-collinear magnetic systems. Thus we expect that these will not apply to collinear AFMs, but will apply to all non-collinear systems, including those that are not AFM.
\bibliography{refs}

\begin{thebibliography}{37}%
\makeatletter
\providecommand \@ifxundefined [1]{%
 \@ifx{#1\undefined}
}%
\providecommand \@ifnum [1]{%
 \ifnum #1\expandafter \@firstoftwo
 \else \expandafter \@secondoftwo
 \fi
}%
\providecommand \@ifx [1]{%
 \ifx #1\expandafter \@firstoftwo
 \else \expandafter \@secondoftwo
 \fi
}%
\providecommand \natexlab [1]{#1}%
\providecommand \enquote  [1]{``#1''}%
\providecommand \bibnamefont  [1]{#1}%
\providecommand \bibfnamefont [1]{#1}%
\providecommand \citenamefont [1]{#1}%
\providecommand \href@noop [0]{\@secondoftwo}%
\providecommand \href [0]{\begingroup \@sanitize@url \@href}%
\providecommand \@href[1]{\@@startlink{#1}\@@href}%
\providecommand \@@href[1]{\endgroup#1\@@endlink}%
\providecommand \@sanitize@url [0]{\catcode `\\12\catcode `\$12\catcode
  `\&12\catcode `\#12\catcode `\^12\catcode `\_12\catcode `\%12\relax}%
\providecommand \@@startlink[1]{}%
\providecommand \@@endlink[0]{}%
\providecommand \url  [0]{\begingroup\@sanitize@url \@url }%
\providecommand \@url [1]{\endgroup\@href {#1}{\urlprefix }}%
\providecommand \urlprefix  [0]{URL }%
\providecommand \Eprint [0]{\href }%
\providecommand \doibase [0]{https://doi.org/}%
\providecommand \selectlanguage [0]{\@gobble}%
\providecommand \bibinfo  [0]{\@secondoftwo}%
\providecommand \bibfield  [0]{\@secondoftwo}%
\providecommand \translation [1]{[#1]}%
\providecommand \BibitemOpen [0]{}%
\providecommand \bibitemStop [0]{}%
\providecommand \bibitemNoStop [0]{.\EOS\space}%
\providecommand \EOS [0]{\spacefactor3000\relax}%
\providecommand \BibitemShut  [1]{\csname bibitem#1\endcsname}%
\let\auto@bib@innerbib\@empty
\bibitem [{\citenamefont {Brataas}\ \emph {et~al.}(2012)\citenamefont
  {Brataas}, \citenamefont {Kent},\ and\ \citenamefont {Ohno}}]{Brataas2012}%
  \BibitemOpen
  \bibfield  {author} {\bibinfo {author} {\bibfnamefont {A.}~\bibnamefont
  {Brataas}}, \bibinfo {author} {\bibfnamefont {A.~D.}\ \bibnamefont {Kent}},\
  and\ \bibinfo {author} {\bibfnamefont {H.}~\bibnamefont {Ohno}},\ }\bibfield
  {title} {\bibinfo {title} {Current-induced torques in magnetic materials},\
  }\href {http://dx.doi.org/10.1038/nmat3311} {\bibfield  {journal} {\bibinfo
  {journal} {Nature Materials}\ }\textbf {\bibinfo {volume} {11}},\ \bibinfo
  {pages} {372 EP } (\bibinfo {year} {2012})},\ \bibinfo {note} {review
  Article}\BibitemShut {NoStop}%
\bibitem [{\citenamefont {Ralph}\ and\ \citenamefont
  {Stiles}(2008)}]{Ralph2008}%
  \BibitemOpen
  \bibfield  {author} {\bibinfo {author} {\bibfnamefont {D.}~\bibnamefont
  {Ralph}}\ and\ \bibinfo {author} {\bibfnamefont {M.}~\bibnamefont {Stiles}},\
  }\bibfield  {title} {\bibinfo {title} {Spin transfer torques},\ }\href
  {https://doi.org/http://dx.doi.org/10.1016/j.jmmm.2007.12.019} {\bibfield
  {journal} {\bibinfo  {journal} {J. Magn. Magn. Matter.}\ }\textbf {\bibinfo
  {volume} {320}},\ \bibinfo {pages} {1190 } (\bibinfo {year}
  {2008})}\BibitemShut {NoStop}%
\bibitem [{\citenamefont {Baibich}\ \emph {et~al.}(1988)\citenamefont
  {Baibich}, \citenamefont {Broto}, \citenamefont {Fert}, \citenamefont
  {Van~Dau}, \citenamefont {Petroff}, \citenamefont {Etienne}, \citenamefont
  {Creuzet}, \citenamefont {Friederich},\ and\ \citenamefont
  {Chazelas}}]{Baibich1988}%
  \BibitemOpen
  \bibfield  {author} {\bibinfo {author} {\bibfnamefont {M.~N.}\ \bibnamefont
  {Baibich}}, \bibinfo {author} {\bibfnamefont {J.~M.}\ \bibnamefont {Broto}},
  \bibinfo {author} {\bibfnamefont {A.}~\bibnamefont {Fert}}, \bibinfo {author}
  {\bibfnamefont {F.~N.}\ \bibnamefont {Van~Dau}}, \bibinfo {author}
  {\bibfnamefont {F.}~\bibnamefont {Petroff}}, \bibinfo {author} {\bibfnamefont
  {P.}~\bibnamefont {Etienne}}, \bibinfo {author} {\bibfnamefont
  {G.}~\bibnamefont {Creuzet}}, \bibinfo {author} {\bibfnamefont
  {A.}~\bibnamefont {Friederich}},\ and\ \bibinfo {author} {\bibfnamefont
  {J.}~\bibnamefont {Chazelas}},\ }\bibfield  {title} {\bibinfo {title} {{Giant
  Magnetoresistance of (001)Fe/(001)Cr Magnetic Superlattices}},\ }\href
  {https://doi.org/10.1103/PhysRevLett.61.2472} {\bibfield  {journal} {\bibinfo
   {journal} {Phys. Rev. Lett.}\ }\textbf {\bibinfo {volume} {61}},\ \bibinfo
  {pages} {2472} (\bibinfo {year} {1988})}\BibitemShut {NoStop}%
\bibitem [{\citenamefont {Binasch}\ \emph {et~al.}(1989)\citenamefont
  {Binasch}, \citenamefont {Gr\"unberg}, \citenamefont {Saurenbach},\ and\
  \citenamefont {Zinn}}]{Binasch1989}%
  \BibitemOpen
  \bibfield  {author} {\bibinfo {author} {\bibfnamefont {G.}~\bibnamefont
  {Binasch}}, \bibinfo {author} {\bibfnamefont {P.}~\bibnamefont {Gr\"unberg}},
  \bibinfo {author} {\bibfnamefont {F.}~\bibnamefont {Saurenbach}},\ and\
  \bibinfo {author} {\bibfnamefont {W.}~\bibnamefont {Zinn}},\ }\bibfield
  {title} {\bibinfo {title} {Enhanced magnetoresistance in layered magnetic
  structures with antiferromagnetic interlayer exchange},\ }\href
  {https://doi.org/10.1103/PhysRevB.39.4828} {\bibfield  {journal} {\bibinfo
  {journal} {Phys. Rev. B}\ }\textbf {\bibinfo {volume} {39}},\ \bibinfo
  {pages} {4828} (\bibinfo {year} {1989})}\BibitemShut {NoStop}%
\bibitem [{\citenamefont {Julliere}(1975)}]{JULLIERE1975225}%
  \BibitemOpen
  \bibfield  {author} {\bibinfo {author} {\bibfnamefont {M.}~\bibnamefont
  {Julliere}},\ }\bibfield  {title} {\bibinfo {title} {Tunneling between
  ferromagnetic films},\ }\href
  {https://doi.org/http://dx.doi.org/10.1016/0375-9601(75)90174-7} {\bibfield
  {journal} {\bibinfo  {journal} {Physics Letters A}\ }\textbf {\bibinfo
  {volume} {54}},\ \bibinfo {pages} {225 } (\bibinfo {year}
  {1975})}\BibitemShut {NoStop}%
\bibitem [{\citenamefont {Apalkov}\ \emph {et~al.}(2016)\citenamefont
  {Apalkov}, \citenamefont {Dieny},\ and\ \citenamefont
  {Slaughter}}]{Apalkov2016}%
  \BibitemOpen
  \bibfield  {author} {\bibinfo {author} {\bibfnamefont {D.}~\bibnamefont
  {Apalkov}}, \bibinfo {author} {\bibfnamefont {B.}~\bibnamefont {Dieny}},\
  and\ \bibinfo {author} {\bibfnamefont {J.~M.}\ \bibnamefont {Slaughter}},\
  }\bibfield  {title} {\bibinfo {title} {Magnetoresistive random access
  memory},\ }\href {https://doi.org/10.1109/JPROC.2016.2590142} {\bibfield
  {journal} {\bibinfo  {journal} {Proceedings of the IEEE}\ }\textbf {\bibinfo
  {volume} {104}},\ \bibinfo {pages} {1796} (\bibinfo {year}
  {2016})}\BibitemShut {NoStop}%
\bibitem [{\citenamefont {Khvalkovskiy}\ \emph {et~al.}(2013)\citenamefont
  {Khvalkovskiy}, \citenamefont {Apalkov}, \citenamefont {Watts}, \citenamefont
  {Chepulskii}, \citenamefont {Beach}, \citenamefont {Ong}, \citenamefont
  {Tang}, \citenamefont {Driskill-Smith}, \citenamefont {Butler}, \citenamefont
  {Visscher}, \citenamefont {Lottis}, \citenamefont {Chen}, \citenamefont
  {Nikitin},\ and\ \citenamefont {Krounbi}}]{Khvalkovskiy2013}%
  \BibitemOpen
  \bibfield  {author} {\bibinfo {author} {\bibfnamefont {A.~V.}\ \bibnamefont
  {Khvalkovskiy}}, \bibinfo {author} {\bibfnamefont {D.}~\bibnamefont
  {Apalkov}}, \bibinfo {author} {\bibfnamefont {S.}~\bibnamefont {Watts}},
  \bibinfo {author} {\bibfnamefont {R.}~\bibnamefont {Chepulskii}}, \bibinfo
  {author} {\bibfnamefont {R.~S.}\ \bibnamefont {Beach}}, \bibinfo {author}
  {\bibfnamefont {A.}~\bibnamefont {Ong}}, \bibinfo {author} {\bibfnamefont
  {X.}~\bibnamefont {Tang}}, \bibinfo {author} {\bibfnamefont {A.}~\bibnamefont
  {Driskill-Smith}}, \bibinfo {author} {\bibfnamefont {W.~H.}\ \bibnamefont
  {Butler}}, \bibinfo {author} {\bibfnamefont {P.~B.}\ \bibnamefont
  {Visscher}}, \bibinfo {author} {\bibfnamefont {D.}~\bibnamefont {Lottis}},
  \bibinfo {author} {\bibfnamefont {E.}~\bibnamefont {Chen}}, \bibinfo {author}
  {\bibfnamefont {V.}~\bibnamefont {Nikitin}},\ and\ \bibinfo {author}
  {\bibfnamefont {M.}~\bibnamefont {Krounbi}},\ }\bibfield  {title} {\bibinfo
  {title} {Basic principles of stt-mram cell operation in memory arrays},\
  }\href {http://stacks.iop.org/0022-3727/46/i=7/a=074001} {\bibfield
  {journal} {\bibinfo  {journal} {Journal of Physics D: Applied Physics}\
  }\textbf {\bibinfo {volume} {46}},\ \bibinfo {pages} {074001} (\bibinfo
  {year} {2013})}\BibitemShut {NoStop}%
\bibitem [{\citenamefont {{Jungwirth}}\ \emph {et~al.}(2016)\citenamefont
  {{Jungwirth}}, \citenamefont {{Marti}}, \citenamefont {{Wadley}},\ and\
  \citenamefont {{Wunderlich}}}]{Jungwirth2015}%
  \BibitemOpen
  \bibfield  {author} {\bibinfo {author} {\bibfnamefont {T.}~\bibnamefont
  {{Jungwirth}}}, \bibinfo {author} {\bibfnamefont {X.}~\bibnamefont
  {{Marti}}}, \bibinfo {author} {\bibfnamefont {P.}~\bibnamefont {{Wadley}}},\
  and\ \bibinfo {author} {\bibfnamefont {J.}~\bibnamefont {{Wunderlich}}},\
  }\bibfield  {title} {\bibinfo {title} {{Antiferromagnetic spintronics}},\
  }\href {https://doi.org/10.1038/nnano.2016.18} {\bibfield  {journal}
  {\bibinfo  {journal} {Nature Nanotechnology}\ }\textbf {\bibinfo {volume}
  {11}},\ \bibinfo {pages} {231} (\bibinfo {year} {2016})}\BibitemShut
  {NoStop}%
\bibitem [{\citenamefont {Baltz}\ \emph {et~al.}(2018)\citenamefont {Baltz},
  \citenamefont {Manchon}, \citenamefont {Tsoi}, \citenamefont {Moriyama},
  \citenamefont {Ono},\ and\ \citenamefont {Tserkovnyak}}]{Baltz2018}%
  \BibitemOpen
  \bibfield  {author} {\bibinfo {author} {\bibfnamefont {V.}~\bibnamefont
  {Baltz}}, \bibinfo {author} {\bibfnamefont {A.}~\bibnamefont {Manchon}},
  \bibinfo {author} {\bibfnamefont {M.}~\bibnamefont {Tsoi}}, \bibinfo {author}
  {\bibfnamefont {T.}~\bibnamefont {Moriyama}}, \bibinfo {author}
  {\bibfnamefont {T.}~\bibnamefont {Ono}},\ and\ \bibinfo {author}
  {\bibfnamefont {Y.}~\bibnamefont {Tserkovnyak}},\ }\bibfield  {title}
  {\bibinfo {title} {Antiferromagnetic spintronics},\ }\href
  {https://doi.org/10.1103/RevModPhys.90.015005} {\bibfield  {journal}
  {\bibinfo  {journal} {Rev. Mod. Phys.}\ }\textbf {\bibinfo {volume} {90}},\
  \bibinfo {pages} {015005} (\bibinfo {year} {2018})}\BibitemShut {NoStop}%
\bibitem [{\citenamefont {Jungwirth}\ \emph {et~al.}(2018)\citenamefont
  {Jungwirth}, \citenamefont {Sinova}, \citenamefont {Manchon}, \citenamefont
  {Marti}, \citenamefont {Wunderlich},\ and\ \citenamefont
  {Felser}}]{Jungwirth2018}%
  \BibitemOpen
  \bibfield  {author} {\bibinfo {author} {\bibfnamefont {T.}~\bibnamefont
  {Jungwirth}}, \bibinfo {author} {\bibfnamefont {J.}~\bibnamefont {Sinova}},
  \bibinfo {author} {\bibfnamefont {A.}~\bibnamefont {Manchon}}, \bibinfo
  {author} {\bibfnamefont {X.}~\bibnamefont {Marti}}, \bibinfo {author}
  {\bibfnamefont {J.}~\bibnamefont {Wunderlich}},\ and\ \bibinfo {author}
  {\bibfnamefont {C.}~\bibnamefont {Felser}},\ }\bibfield  {title} {\bibinfo
  {title} {The multiple directions of antiferromagnetic spintronics},\ }\href
  {https://doi.org/10.1038/s41567-018-0063-6} {\bibfield  {journal} {\bibinfo
  {journal} {Nature Physics}\ }\textbf {\bibinfo {volume} {14}},\ \bibinfo
  {pages} {200} (\bibinfo {year} {2018})}\BibitemShut {NoStop}%
\bibitem [{\citenamefont {{{\v Z}elezn{\'y}}}\ \emph
  {et~al.}(2018)\citenamefont {{{\v Z}elezn{\'y}}}, \citenamefont {{Wadley}},
  \citenamefont {{Olejn{\'i}k}}, \citenamefont {{Hoffmann}},\ and\
  \citenamefont {{Ohno}}}]{Zelezny2018}%
  \BibitemOpen
  \bibfield  {author} {\bibinfo {author} {\bibfnamefont {J.}~\bibnamefont {{{\v
  Z}elezn{\'y}}}}, \bibinfo {author} {\bibfnamefont {P.}~\bibnamefont
  {{Wadley}}}, \bibinfo {author} {\bibfnamefont {K.}~\bibnamefont
  {{Olejn{\'i}k}}}, \bibinfo {author} {\bibfnamefont {A.}~\bibnamefont
  {{Hoffmann}}},\ and\ \bibinfo {author} {\bibfnamefont {H.}~\bibnamefont
  {{Ohno}}},\ }\bibfield  {title} {\bibinfo {title} {{Spin-transport,
  spin-torque and memory in antiferromagnetic devices}},\ }\href
  {https://doi.org/10.1038/s41567-018-0062-7} {\bibfield  {journal} {\bibinfo
  {journal} {Nature Phys.}\ }\textbf {\bibinfo {volume} {14}},\ \bibinfo
  {pages} {220} (\bibinfo {year} {2018})}\BibitemShut {NoStop}%
\bibitem [{\citenamefont {{Eerenstein}}\ \emph {et~al.}(2006)\citenamefont
  {{Eerenstein}}, \citenamefont {{Mathur}},\ and\ \citenamefont
  {{Scott}}}]{Eerenstein2006}%
  \BibitemOpen
  \bibfield  {author} {\bibinfo {author} {\bibfnamefont {W.}~\bibnamefont
  {{Eerenstein}}}, \bibinfo {author} {\bibfnamefont {N.~D.}\ \bibnamefont
  {{Mathur}}},\ and\ \bibinfo {author} {\bibfnamefont {J.~F.}\ \bibnamefont
  {{Scott}}},\ }\bibfield  {title} {\bibinfo {title} {{Multiferroic and
  magnetoelectric materials}},\ }\href {https://doi.org/10.1038/nature05023}
  {\bibfield  {journal} {\bibinfo  {journal} {Nature}\ }\textbf {\bibinfo
  {volume} {442}},\ \bibinfo {pages} {759} (\bibinfo {year}
  {2006})}\BibitemShut {NoStop}%
\bibitem [{\citenamefont {{Lu}}\ \emph {et~al.}(2015)\citenamefont {{Lu}},
  \citenamefont {{Wang}}, \citenamefont {{Wu}}, \citenamefont {{Wu}},
  \citenamefont {{Zhao}}, \citenamefont {{Zeng}}, \citenamefont {{Luo}},
  \citenamefont {{Wu}}, \citenamefont {{Bao}}, \citenamefont {{Zhang}},
  \citenamefont {{Huang}}, \citenamefont {{Huang}},\ and\ \citenamefont
  {{Chen}}}]{Lu2015}%
  \BibitemOpen
  \bibfield  {author} {\bibinfo {author} {\bibfnamefont {X.~F.}\ \bibnamefont
  {{Lu}}}, \bibinfo {author} {\bibfnamefont {N.~Z.}\ \bibnamefont {{Wang}}},
  \bibinfo {author} {\bibfnamefont {H.}~\bibnamefont {{Wu}}}, \bibinfo {author}
  {\bibfnamefont {Y.~P.}\ \bibnamefont {{Wu}}}, \bibinfo {author}
  {\bibfnamefont {D.}~\bibnamefont {{Zhao}}}, \bibinfo {author} {\bibfnamefont
  {X.~Z.}\ \bibnamefont {{Zeng}}}, \bibinfo {author} {\bibfnamefont {X.~G.}\
  \bibnamefont {{Luo}}}, \bibinfo {author} {\bibfnamefont {T.}~\bibnamefont
  {{Wu}}}, \bibinfo {author} {\bibfnamefont {W.}~\bibnamefont {{Bao}}},
  \bibinfo {author} {\bibfnamefont {G.~H.}\ \bibnamefont {{Zhang}}}, \bibinfo
  {author} {\bibfnamefont {F.~Q.}\ \bibnamefont {{Huang}}}, \bibinfo {author}
  {\bibfnamefont {Q.~Z.}\ \bibnamefont {{Huang}}},\ and\ \bibinfo {author}
  {\bibfnamefont {X.~H.}\ \bibnamefont {{Chen}}},\ }\bibfield  {title}
  {\bibinfo {title} {{Coexistence of superconductivity and antiferromagnetism
  in (Li$_{0.8}$Fe$_{0.2}$)OHFeSe}},\ }\href {https://doi.org/10.1038/nmat4155}
  {\bibfield  {journal} {\bibinfo  {journal} {Nature Materials}\ }\textbf
  {\bibinfo {volume} {14}},\ \bibinfo {pages} {325} (\bibinfo {year}
  {2015})}\BibitemShut {NoStop}%
\bibitem [{\citenamefont {N\'u\~nez}\ \emph {et~al.}(2006)\citenamefont
  {N\'u\~nez}, \citenamefont {Duine}, \citenamefont {Haney},\ and\
  \citenamefont {MacDonald}}]{Nunez2006}%
  \BibitemOpen
  \bibfield  {author} {\bibinfo {author} {\bibfnamefont {A.~S.}\ \bibnamefont
  {N\'u\~nez}}, \bibinfo {author} {\bibfnamefont {R.~A.}\ \bibnamefont
  {Duine}}, \bibinfo {author} {\bibfnamefont {P.}~\bibnamefont {Haney}},\ and\
  \bibinfo {author} {\bibfnamefont {A.~H.}\ \bibnamefont {MacDonald}},\
  }\bibfield  {title} {\bibinfo {title} {Theory of spin torques and giant
  magnetoresistance in antiferromagnetic metals},\ }\href
  {https://doi.org/10.1103/PhysRevB.73.214426} {\bibfield  {journal} {\bibinfo
  {journal} {Phys. Rev. B}\ }\textbf {\bibinfo {volume} {73}},\ \bibinfo
  {pages} {214426} (\bibinfo {year} {2006})}\BibitemShut {NoStop}%
\bibitem [{\citenamefont {Haney}\ \emph {et~al.}(2007)\citenamefont {Haney},
  \citenamefont {Waldron}, \citenamefont {Duine}, \citenamefont
  {N{\'{u}}{\~{n}}ez}, \citenamefont {Guo},\ and\ \citenamefont
  {MacDonald}}]{Haney2007}%
  \BibitemOpen
  \bibfield  {author} {\bibinfo {author} {\bibfnamefont {P.~M.}\ \bibnamefont
  {Haney}}, \bibinfo {author} {\bibfnamefont {D.}~\bibnamefont {Waldron}},
  \bibinfo {author} {\bibfnamefont {R.~A.}\ \bibnamefont {Duine}}, \bibinfo
  {author} {\bibfnamefont {A.}~\bibnamefont {N{\'{u}}{\~{n}}ez}}, \bibinfo
  {author} {\bibfnamefont {H.}~\bibnamefont {Guo}},\ and\ \bibinfo {author}
  {\bibfnamefont {a.~H.}\ \bibnamefont {MacDonald}},\ }\bibfield  {title}
  {\bibinfo {title} {{Ab initio giant magnetoresistance and current-induced
  torques in Cr/Au/Cr multilayers}},\ }\href
  {https://doi.org/10.1103/PhysRevB.75.174428} {\bibfield  {journal} {\bibinfo
  {journal} {Phys. Rev. B}\ }\textbf {\bibinfo {volume} {75}},\ \bibinfo
  {pages} {174428(1)} (\bibinfo {year} {2007})}\BibitemShut {NoStop}%
\bibitem [{\citenamefont {Merodio}\ \emph {et~al.}(2014)\citenamefont
  {Merodio}, \citenamefont {Kalitsov}, \citenamefont {B{\,e}a}, \citenamefont
  {Baltz},\ and\ \citenamefont {Chshiev}}]{Merodio2014}%
  \BibitemOpen
  \bibfield  {author} {\bibinfo {author} {\bibfnamefont {P.}~\bibnamefont
  {Merodio}}, \bibinfo {author} {\bibfnamefont {A.}~\bibnamefont {Kalitsov}},
  \bibinfo {author} {\bibfnamefont {H.}~\bibnamefont {B{\,e}a}}, \bibinfo
  {author} {\bibfnamefont {V.}~\bibnamefont {Baltz}},\ and\ \bibinfo {author}
  {\bibfnamefont {M.}~\bibnamefont {Chshiev}},\ }\bibfield  {title} {\bibinfo
  {title} {Spin-dependent transport in antiferromagnetic tunnel junctions},\
  }\href {https://doi.org/10.1063/1.4896291} {\bibfield  {journal} {\bibinfo
  {journal} {Appl. Phys. Lett.}\ }\textbf {\bibinfo {volume} {105}},\ \bibinfo
  {pages} {122403} (\bibinfo {year} {2014})}\BibitemShut {NoStop}%
\bibitem [{\citenamefont {Stamenova}\ \emph {et~al.}(2017)\citenamefont
  {Stamenova}, \citenamefont {Mohebbi}, \citenamefont {Seyed-Yazdi},
  \citenamefont {Rungger},\ and\ \citenamefont {Sanvito}}]{Stamenova2017}%
  \BibitemOpen
  \bibfield  {author} {\bibinfo {author} {\bibfnamefont {M.}~\bibnamefont
  {Stamenova}}, \bibinfo {author} {\bibfnamefont {R.}~\bibnamefont {Mohebbi}},
  \bibinfo {author} {\bibfnamefont {J.}~\bibnamefont {Seyed-Yazdi}}, \bibinfo
  {author} {\bibfnamefont {I.}~\bibnamefont {Rungger}},\ and\ \bibinfo {author}
  {\bibfnamefont {S.}~\bibnamefont {Sanvito}},\ }\bibfield  {title} {\bibinfo
  {title} {First-principles spin-transfer torque in
  {$\mathrm{CuMnAs}|\mathrm{GaP}|\mathrm{CuMnAs}$} junctions},\ }\href
  {https://doi.org/10.1103/PhysRevB.95.060403} {\bibfield  {journal} {\bibinfo
  {journal} {Phys. Rev. B}\ }\textbf {\bibinfo {volume} {95}},\ \bibinfo
  {pages} {060403} (\bibinfo {year} {2017})}\BibitemShut {NoStop}%
\bibitem [{\citenamefont {Duine}\ \emph {et~al.}(2007)\citenamefont {Duine},
  \citenamefont {Haney}, \citenamefont {N{\'{u}}{\~{n}}ez},\ and\ \citenamefont
  {MacDonald}}]{Duine2007}%
  \BibitemOpen
  \bibfield  {author} {\bibinfo {author} {\bibfnamefont {R.~A.}\ \bibnamefont
  {Duine}}, \bibinfo {author} {\bibfnamefont {P.~M.}\ \bibnamefont {Haney}},
  \bibinfo {author} {\bibfnamefont {A.}~\bibnamefont {N{\'{u}}{\~{n}}ez}},\
  and\ \bibinfo {author} {\bibfnamefont {A.}~\bibnamefont {MacDonald}},\
  }\bibfield  {title} {\bibinfo {title} {{Inelastic scattering in ferromagnetic
  and antiferromagnetic spin valves}},\ }\href
  {https://doi.org/10.1103/PhysRevB.75.014433} {\bibfield  {journal} {\bibinfo
  {journal} {Phys. Rev. B}\ }\textbf {\bibinfo {volume} {75}},\ \bibinfo
  {pages} {014433} (\bibinfo {year} {2007})}\BibitemShut {NoStop}%
\bibitem [{\citenamefont {Saidaoui}\ \emph {et~al.}(2014)\citenamefont
  {Saidaoui}, \citenamefont {Manchon},\ and\ \citenamefont
  {Waintal}}]{Saidaoui2014}%
  \BibitemOpen
  \bibfield  {author} {\bibinfo {author} {\bibfnamefont {H.~B.~M.}\
  \bibnamefont {Saidaoui}}, \bibinfo {author} {\bibfnamefont {A.}~\bibnamefont
  {Manchon}},\ and\ \bibinfo {author} {\bibfnamefont {X.}~\bibnamefont
  {Waintal}},\ }\bibfield  {title} {\bibinfo {title} {Spin transfer torque in
  antiferromagnetic spin valves: From clean to disordered regimes},\ }\href
  {https://doi.org/10.1103/PhysRevB.89.174430} {\bibfield  {journal} {\bibinfo
  {journal} {Phys. Rev. B}\ }\textbf {\bibinfo {volume} {89}},\ \bibinfo
  {pages} {174430} (\bibinfo {year} {2014})}\BibitemShut {NoStop}%
\bibitem [{\citenamefont {Manchon}(2017)}]{Manchon2017}%
  \BibitemOpen
  \bibfield  {author} {\bibinfo {author} {\bibfnamefont {A.}~\bibnamefont
  {Manchon}},\ }\bibfield  {title} {\bibinfo {title} {Spin diffusion and
  torques in disordered antiferromagnets},\ }\href
  {http://stacks.iop.org/0953-8984/29/i=10/a=104002} {\bibfield  {journal}
  {\bibinfo  {journal} {Journal of Physics: Condensed Matter}\ }\textbf
  {\bibinfo {volume} {29}},\ \bibinfo {pages} {104002} (\bibinfo {year}
  {2017})}\BibitemShut {NoStop}%
\bibitem [{\citenamefont {Saidaoui}\ \emph {et~al.}(2017)\citenamefont
  {Saidaoui}, \citenamefont {Waintal},\ and\ \citenamefont
  {Manchon}}]{Saidaoui2016a}%
  \BibitemOpen
  \bibfield  {author} {\bibinfo {author} {\bibfnamefont {H.~B.~M.}\
  \bibnamefont {Saidaoui}}, \bibinfo {author} {\bibfnamefont {X.}~\bibnamefont
  {Waintal}},\ and\ \bibinfo {author} {\bibfnamefont {A.}~\bibnamefont
  {Manchon}},\ }\bibfield  {title} {\bibinfo {title} {Robust spin transfer
  torque in antiferromagnetic tunnel junctions},\ }\href
  {https://doi.org/10.1103/PhysRevB.95.134424} {\bibfield  {journal} {\bibinfo
  {journal} {Phys. Rev. B}\ }\textbf {\bibinfo {volume} {95}},\ \bibinfo
  {pages} {134424} (\bibinfo {year} {2017})}\BibitemShut {NoStop}%
\bibitem [{\citenamefont {{\v{Z}}elezn{\'{y}}}\ \emph
  {et~al.}(2017)\citenamefont {{\v{Z}}elezn{\'{y}}}, \citenamefont {Zhang},
  \citenamefont {Felser},\ and\ \citenamefont {Yan}}]{Zelezny2017b}%
  \BibitemOpen
  \bibfield  {author} {\bibinfo {author} {\bibfnamefont {J.}~\bibnamefont
  {{\v{Z}}elezn{\'{y}}}}, \bibinfo {author} {\bibfnamefont {Y.}~\bibnamefont
  {Zhang}}, \bibinfo {author} {\bibfnamefont {C.}~\bibnamefont {Felser}},\ and\
  \bibinfo {author} {\bibfnamefont {B.}~\bibnamefont {Yan}},\ }\bibfield
  {title} {\bibinfo {title} {Spin-polarized current in noncollinear
  antiferromagnets},\ }\href {https://doi.org/10.1103/PhysRevLett.119.187204}
  {\bibfield  {journal} {\bibinfo  {journal} {Phys. Rev. Lett.}\ }\textbf
  {\bibinfo {volume} {119}},\ \bibinfo {pages} {187204} (\bibinfo {year}
  {2017})}\BibitemShut {NoStop}%
\bibitem [{\citenamefont {Naka}\ \emph {et~al.}(2019)\citenamefont {Naka},
  \citenamefont {Hayami}, \citenamefont {Kusunose}, \citenamefont {Yanagi},
  \citenamefont {Motome},\ and\ \citenamefont {Seo}}]{Naka2019}%
  \BibitemOpen
  \bibfield  {author} {\bibinfo {author} {\bibfnamefont {M.}~\bibnamefont
  {Naka}}, \bibinfo {author} {\bibfnamefont {S.}~\bibnamefont {Hayami}},
  \bibinfo {author} {\bibfnamefont {H.}~\bibnamefont {Kusunose}}, \bibinfo
  {author} {\bibfnamefont {Y.}~\bibnamefont {Yanagi}}, \bibinfo {author}
  {\bibfnamefont {Y.}~\bibnamefont {Motome}},\ and\ \bibinfo {author}
  {\bibfnamefont {H.}~\bibnamefont {Seo}},\ }\bibfield  {title} {\bibinfo
  {title} {Spin current generation in organic antiferromagnets},\ }\href@noop
  {} {\bibfield  {journal} {\bibinfo  {journal} {Nature communications}\
  }\textbf {\bibinfo {volume} {10}},\ \bibinfo {pages} {1} (\bibinfo {year}
  {2019})}\BibitemShut {NoStop}%
\bibitem [{\citenamefont {Gonz\'alez-Hern\'andez}\ \emph
  {et~al.}(2021)\citenamefont {Gonz\'alez-Hern\'andez}, \citenamefont
  {\ifmmode~\check{S}\else \v{S}\fi{}mejkal}, \citenamefont {V\'yborn\'y},
  \citenamefont {Yahagi}, \citenamefont {Sinova}, \citenamefont {Jungwirth},\
  and\ \citenamefont {\ifmmode~\check{Z}\else
  \v{Z}\fi{}elezn\'y}}]{Hernandez2021}%
  \BibitemOpen
  \bibfield  {author} {\bibinfo {author} {\bibfnamefont {R.}~\bibnamefont
  {Gonz\'alez-Hern\'andez}}, \bibinfo {author} {\bibfnamefont {L.}~\bibnamefont
  {\ifmmode~\check{S}\else \v{S}\fi{}mejkal}}, \bibinfo {author} {\bibfnamefont
  {K.}~\bibnamefont {V\'yborn\'y}}, \bibinfo {author} {\bibfnamefont
  {Y.}~\bibnamefont {Yahagi}}, \bibinfo {author} {\bibfnamefont
  {J.}~\bibnamefont {Sinova}}, \bibinfo {author} {\bibfnamefont {T.~c.~v.}\
  \bibnamefont {Jungwirth}},\ and\ \bibinfo {author} {\bibfnamefont
  {J.}~\bibnamefont {\ifmmode~\check{Z}\else \v{Z}\fi{}elezn\'y}},\ }\bibfield
  {title} {\bibinfo {title} {Efficient electrical spin splitter based on
  nonrelativistic collinear antiferromagnetism},\ }\href
  {https://doi.org/10.1103/PhysRevLett.126.127701} {\bibfield  {journal}
  {\bibinfo  {journal} {Phys. Rev. Lett.}\ }\textbf {\bibinfo {volume} {126}},\
  \bibinfo {pages} {127701} (\bibinfo {year} {2021})}\BibitemShut {NoStop}%
\bibitem [{\citenamefont {Ahn}\ \emph {et~al.}(2019)\citenamefont {Ahn},
  \citenamefont {Hariki}, \citenamefont {Lee},\ and\ \citenamefont
  {Kune\ifmmode~\check{s}\else \v{s}\fi{}}}]{Ahn2019}%
  \BibitemOpen
  \bibfield  {author} {\bibinfo {author} {\bibfnamefont {K.-H.}\ \bibnamefont
  {Ahn}}, \bibinfo {author} {\bibfnamefont {A.}~\bibnamefont {Hariki}},
  \bibinfo {author} {\bibfnamefont {K.-W.}\ \bibnamefont {Lee}},\ and\ \bibinfo
  {author} {\bibfnamefont {J.}~\bibnamefont {Kune\ifmmode~\check{s}\else
  \v{s}\fi{}}},\ }\bibfield  {title} {\bibinfo {title} {Antiferromagnetism in
  ${\mathrm{ruo}}_{2}$ as $d$-wave pomeranchuk instability},\ }\href
  {https://doi.org/10.1103/PhysRevB.99.184432} {\bibfield  {journal} {\bibinfo
  {journal} {Phys. Rev. B}\ }\textbf {\bibinfo {volume} {99}},\ \bibinfo
  {pages} {184432} (\bibinfo {year} {2019})}\BibitemShut {NoStop}%
\bibitem [{\citenamefont {Hayami}\ \emph {et~al.}(2019)\citenamefont {Hayami},
  \citenamefont {Yanagi},\ and\ \citenamefont {Kusunose}}]{Hayami2019}%
  \BibitemOpen
  \bibfield  {author} {\bibinfo {author} {\bibfnamefont {S.}~\bibnamefont
  {Hayami}}, \bibinfo {author} {\bibfnamefont {Y.}~\bibnamefont {Yanagi}},\
  and\ \bibinfo {author} {\bibfnamefont {H.}~\bibnamefont {Kusunose}},\
  }\bibfield  {title} {\bibinfo {title} {{Momentum-Dependent Spin Splitting by
  Collinear Antiferromagnetic Ordering}},\ }\href
  {https://doi.org/10.7566/JPSJ.88.123702} {\bibfield  {journal} {\bibinfo
  {journal} {Journal of the Physical Society of Japan}\ }\textbf {\bibinfo
  {volume} {88}},\ \bibinfo {pages} {123702} (\bibinfo {year}
  {2019})}\BibitemShut {NoStop}%
\bibitem [{\citenamefont {{\v{S}}mejkal}\ \emph {et~al.}(2020)\citenamefont
  {{\v{S}}mejkal}, \citenamefont {Gonz{\'{a}}lez-Hern{\'{a}}ndez},
  \citenamefont {Jungwirth},\ and\ \citenamefont {Sinova}}]{Smejkal2020}%
  \BibitemOpen
  \bibfield  {author} {\bibinfo {author} {\bibfnamefont {L.}~\bibnamefont
  {{\v{S}}mejkal}}, \bibinfo {author} {\bibfnamefont {R.}~\bibnamefont
  {Gonz{\'{a}}lez-Hern{\'{a}}ndez}}, \bibinfo {author} {\bibfnamefont
  {T.}~\bibnamefont {Jungwirth}},\ and\ \bibinfo {author} {\bibfnamefont
  {J.}~\bibnamefont {Sinova}},\ }\bibfield  {title} {\bibinfo {title} {{Crystal
  time-reversal symmetry breaking and spontaneous Hall effect in collinear
  antiferromagnets}},\ }\href {https://doi.org/10.1126/sciadv.aaz8809}
  {\bibfield  {journal} {\bibinfo  {journal} {Science Advances}\ }\textbf
  {\bibinfo {volume} {6}},\ \bibinfo {pages} {eaaz8809} (\bibinfo {year}
  {2020})},\ \Eprint {https://arxiv.org/abs/1901.00445} {arXiv:1901.00445}
  \BibitemShut {NoStop}%
\bibitem [{\citenamefont {Yuan}\ \emph {et~al.}(2020)\citenamefont {Yuan},
  \citenamefont {Wang}, \citenamefont {Luo}, \citenamefont {Rashba},\ and\
  \citenamefont {Zunger}}]{Yuan2020}%
  \BibitemOpen
  \bibfield  {author} {\bibinfo {author} {\bibfnamefont {L.-D.}\ \bibnamefont
  {Yuan}}, \bibinfo {author} {\bibfnamefont {Z.}~\bibnamefont {Wang}}, \bibinfo
  {author} {\bibfnamefont {J.-W.}\ \bibnamefont {Luo}}, \bibinfo {author}
  {\bibfnamefont {E.~I.}\ \bibnamefont {Rashba}},\ and\ \bibinfo {author}
  {\bibfnamefont {A.}~\bibnamefont {Zunger}},\ }\bibfield  {title} {\bibinfo
  {title} {{Giant momentum-dependent spin splitting in centrosymmetric low-
  {\textless}math{\textgreater}
  {\textless}mi{\textgreater}Z{\textless}/mi{\textgreater}
  {\textless}/math{\textgreater} antiferromagnets}},\ }\href
  {https://doi.org/10.1103/PhysRevB.102.014422} {\bibfield  {journal} {\bibinfo
   {journal} {Physical Review B}\ }\textbf {\bibinfo {volume} {102}},\ \bibinfo
  {pages} {014422} (\bibinfo {year} {2020})}\BibitemShut {NoStop}%
\bibitem [{\citenamefont {{\v{S}}mejkal}\ \emph {et~al.}(2021)\citenamefont
  {{\v{S}}mejkal}, \citenamefont {Hellenes}, \citenamefont
  {Gonz{\'a}lez-Hern{\'a}ndez}, \citenamefont {Sinova},\ and\ \citenamefont
  {Jungwirth}}]{Smejkal2021giant}%
  \BibitemOpen
  \bibfield  {author} {\bibinfo {author} {\bibfnamefont {L.}~\bibnamefont
  {{\v{S}}mejkal}}, \bibinfo {author} {\bibfnamefont {A.~B.}\ \bibnamefont
  {Hellenes}}, \bibinfo {author} {\bibfnamefont {R.}~\bibnamefont
  {Gonz{\'a}lez-Hern{\'a}ndez}}, \bibinfo {author} {\bibfnamefont
  {J.}~\bibnamefont {Sinova}},\ and\ \bibinfo {author} {\bibfnamefont
  {T.}~\bibnamefont {Jungwirth}},\ }\bibfield  {title} {\bibinfo {title} {Giant
  and tunneling magnetoresistance effects from anisotropic and valley-dependent
  spin-momentum interactions in antiferromagnets},\ }\href@noop {} {\bibfield
  {journal} {\bibinfo  {journal} {arXiv:2103.12664}\ } (\bibinfo {year}
  {2021})}\BibitemShut {NoStop}%
\bibitem [{\citenamefont {Tomiyoshi}\ and\ \citenamefont
  {Yamaguchi}(1982)}]{Tomiyoshi1982}%
  \BibitemOpen
  \bibfield  {author} {\bibinfo {author} {\bibfnamefont {S.}~\bibnamefont
  {Tomiyoshi}}\ and\ \bibinfo {author} {\bibfnamefont {Y.}~\bibnamefont
  {Yamaguchi}},\ }\bibfield  {title} {\bibinfo {title} {{Magnetic Structure and
  Weak Ferromagnetism of Mn3Sn Studied by Polarized Neutron Diffraction}},\
  }\href {https://doi.org/10.1143/JPSJ.51.2478} {\bibfield  {journal} {\bibinfo
   {journal} {Journal of the Physical Society of Japan}\ }\textbf {\bibinfo
  {volume} {51}},\ \bibinfo {pages} {2478} (\bibinfo {year}
  {1982})}\BibitemShut {NoStop}%
\bibitem [{\citenamefont {Yamaoka}(1974)}]{Yamaoka1974}%
  \BibitemOpen
  \bibfield  {author} {\bibinfo {author} {\bibfnamefont {T.}~\bibnamefont
  {Yamaoka}},\ }\bibfield  {title} {\bibinfo {title} {Antiferromagnetism in
  $\gamma$-phase mn-ir alloys},\ }\href {https://doi.org/10.1143/JPSJ.36.445}
  {\bibfield  {journal} {\bibinfo  {journal} {Journal of the Physical Society
  of Japan}\ }\textbf {\bibinfo {volume} {36}},\ \bibinfo {pages} {445}
  (\bibinfo {year} {1974})}\BibitemShut {NoStop}%
\bibitem [{\citenamefont {Chen}\ \emph {et~al.}(2014)\citenamefont {Chen},
  \citenamefont {Niu},\ and\ \citenamefont {MacDonald}}]{Chen2014}%
  \BibitemOpen
  \bibfield  {author} {\bibinfo {author} {\bibfnamefont {H.}~\bibnamefont
  {Chen}}, \bibinfo {author} {\bibfnamefont {Q.}~\bibnamefont {Niu}},\ and\
  \bibinfo {author} {\bibfnamefont {A.~H.}\ \bibnamefont {MacDonald}},\
  }\bibfield  {title} {\bibinfo {title} {Anomalous hall effect arising from
  noncollinear antiferromagnetism},\ }\href
  {https://doi.org/10.1103/PhysRevLett.112.017205} {\bibfield  {journal}
  {\bibinfo  {journal} {Phys. Rev. Lett.}\ }\textbf {\bibinfo {volume} {112}},\
  \bibinfo {pages} {017205} (\bibinfo {year} {2014})}\BibitemShut {NoStop}%
\bibitem [{\citenamefont {Zhang}\ \emph {et~al.}(2018)\citenamefont {Zhang},
  \citenamefont {{{\v Z}elezn{\'y}}}, \citenamefont {Sun}, \citenamefont
  {van~den Brink},\ and\ \citenamefont {Yan}}]{Zhang2018}%
  \BibitemOpen
  \bibfield  {author} {\bibinfo {author} {\bibfnamefont {Y.}~\bibnamefont
  {Zhang}}, \bibinfo {author} {\bibfnamefont {J.}~\bibnamefont {{{\v
  Z}elezn{\'y}}}}, \bibinfo {author} {\bibfnamefont {Y.}~\bibnamefont {Sun}},
  \bibinfo {author} {\bibfnamefont {J.}~\bibnamefont {van~den Brink}},\ and\
  \bibinfo {author} {\bibfnamefont {B.}~\bibnamefont {Yan}},\ }\bibfield
  {title} {\bibinfo {title} {{Spin Hall effect emerging from a noncollinear
  magnetic lattice without spin-orbit coupling}},\ }\href
  {http://stacks.iop.org/1367-2630/20/i=7/a=073028} {\bibfield  {journal}
  {\bibinfo  {journal} {New Journal of Physics}\ }\textbf {\bibinfo {volume}
  {20}},\ \bibinfo {pages} {073028} (\bibinfo {year} {2018})}\BibitemShut
  {NoStop}%
\bibitem [{\citenamefont {Hayami}\ \emph {et~al.}(2020)\citenamefont {Hayami},
  \citenamefont {Yanagi},\ and\ \citenamefont {Kusunose}}]{Hayami2020}%
  \BibitemOpen
  \bibfield  {author} {\bibinfo {author} {\bibfnamefont {S.}~\bibnamefont
  {Hayami}}, \bibinfo {author} {\bibfnamefont {Y.}~\bibnamefont {Yanagi}},\
  and\ \bibinfo {author} {\bibfnamefont {H.}~\bibnamefont {Kusunose}},\
  }\bibfield  {title} {\bibinfo {title} {Spontaneous antisymmetric spin
  splitting in noncollinear antiferromagnets without spin-orbit coupling},\
  }\href {https://doi.org/10.1103/PhysRevB.101.220403} {\bibfield  {journal}
  {\bibinfo  {journal} {Phys. Rev. B}\ }\textbf {\bibinfo {volume} {101}},\
  \bibinfo {pages} {220403} (\bibinfo {year} {2020})}\BibitemShut {NoStop}%
\bibitem [{\citenamefont {Groth}\ \emph {et~al.}(2014)\citenamefont {Groth},
  \citenamefont {Wimmer}, \citenamefont {Akhmerov},\ and\ \citenamefont
  {Waintal}}]{Groth2014}%
  \BibitemOpen
  \bibfield  {author} {\bibinfo {author} {\bibfnamefont {C.~W.}\ \bibnamefont
  {Groth}}, \bibinfo {author} {\bibfnamefont {M.}~\bibnamefont {Wimmer}},
  \bibinfo {author} {\bibfnamefont {A.~R.}\ \bibnamefont {Akhmerov}},\ and\
  \bibinfo {author} {\bibfnamefont {X.}~\bibnamefont {Waintal}},\ }\bibfield
  {title} {\bibinfo {title} {Kwant: a software package for quantum transport},\
  }\href {http://stacks.iop.org/1367-2630/16/i=6/a=063065} {\bibfield
  {journal} {\bibinfo  {journal} {New Journal of Physics}\ }\textbf {\bibinfo
  {volume} {16}},\ \bibinfo {pages} {063065} (\bibinfo {year}
  {2014})}\BibitemShut {NoStop}%
\bibitem [{\citenamefont {Gomonay}\ \emph {et~al.}(2012)\citenamefont
  {Gomonay}, \citenamefont {Kunitsyn},\ and\ \citenamefont
  {Loktev}}]{Gomonay2012}%
  \BibitemOpen
  \bibfield  {author} {\bibinfo {author} {\bibfnamefont {H.~V.}\ \bibnamefont
  {Gomonay}}, \bibinfo {author} {\bibfnamefont {R.~V.}\ \bibnamefont
  {Kunitsyn}},\ and\ \bibinfo {author} {\bibfnamefont {V.~M.}\ \bibnamefont
  {Loktev}},\ }\bibfield  {title} {\bibinfo {title} {{Symmetry and the
  macroscopic dynamics of antiferromagnetic materials in the presence of
  spin-polarized current}},\ }\href
  {https://doi.org/10.1103/PhysRevB.85.134446} {\bibfield  {journal} {\bibinfo
  {journal} {Phys. Rev. B}\ }\textbf {\bibinfo {volume} {85}},\ \bibinfo
  {pages} {134446} (\bibinfo {year} {2012})}\BibitemShut {NoStop}%
\bibitem [{\citenamefont {Cheng}\ \emph {et~al.}(2014)\citenamefont {Cheng},
  \citenamefont {Xiao}, \citenamefont {Niu},\ and\ \citenamefont
  {Brataas}}]{Cheng2014c}%
  \BibitemOpen
  \bibfield  {author} {\bibinfo {author} {\bibfnamefont {R.}~\bibnamefont
  {Cheng}}, \bibinfo {author} {\bibfnamefont {J.}~\bibnamefont {Xiao}},
  \bibinfo {author} {\bibfnamefont {Q.}~\bibnamefont {Niu}},\ and\ \bibinfo
  {author} {\bibfnamefont {A.}~\bibnamefont {Brataas}},\ }\bibfield  {title}
  {\bibinfo {title} {{Spin pumping and spin-transfer torques in
  antiferromagnets}},\ }\href {https://doi.org/10.1103/PhysRevLett.113.057601}
  {\bibfield  {journal} {\bibinfo  {journal} {Phys. Rev. Lett.}\ }\textbf
  {\bibinfo {volume} {113}},\ \bibinfo {pages} {057601} (\bibinfo {year}
  {2014})}\BibitemShut {NoStop}%
\end{thebibliography}%


\begin{thebibliography}{7}%
\makeatletter
\providecommand \@ifxundefined [1]{%
 \@ifx{#1\undefined}
}%
\providecommand \@ifnum [1]{%
 \ifnum #1\expandafter \@firstoftwo
 \else \expandafter \@secondoftwo
 \fi
}%
\providecommand \@ifx [1]{%
 \ifx #1\expandafter \@firstoftwo
 \else \expandafter \@secondoftwo
 \fi
}%
\providecommand \natexlab [1]{#1}%
\providecommand \enquote  [1]{``#1''}%
\providecommand \bibnamefont  [1]{#1}%
\providecommand \bibfnamefont [1]{#1}%
\providecommand \citenamefont [1]{#1}%
\providecommand \href@noop [0]{\@secondoftwo}%
\providecommand \href [0]{\begingroup \@sanitize@url \@href}%
\providecommand \@href[1]{\@@startlink{#1}\@@href}%
\providecommand \@@href[1]{\endgroup#1\@@endlink}%
\providecommand \@sanitize@url [0]{\catcode `\\12\catcode `\$12\catcode
  `\&12\catcode `\#12\catcode `\^12\catcode `\_12\catcode `\%12\relax}%
\providecommand \@@startlink[1]{}%
\providecommand \@@endlink[0]{}%
\providecommand \url  [0]{\begingroup\@sanitize@url \@url }%
\providecommand \@url [1]{\endgroup\@href {#1}{\urlprefix }}%
\providecommand \urlprefix  [0]{URL }%
\providecommand \Eprint [0]{\href }%
\providecommand \doibase [0]{https://doi.org/}%
\providecommand \selectlanguage [0]{\@gobble}%
\providecommand \bibinfo  [0]{\@secondoftwo}%
\providecommand \bibfield  [0]{\@secondoftwo}%
\providecommand \translation [1]{[#1]}%
\providecommand \BibitemOpen [0]{}%
\providecommand \bibitemStop [0]{}%
\providecommand \bibitemNoStop [0]{.\EOS\space}%
\providecommand \EOS [0]{\spacefactor3000\relax}%
\providecommand \BibitemShut  [1]{\csname bibitem#1\endcsname}%
\let\auto@bib@innerbib\@empty
\bibitem [{\citenamefont {Gambardella}\ and\ \citenamefont
  {Miron}(2011)}]{Gambardella2011}%
  \BibitemOpen
  \bibfield  {author} {\bibinfo {author} {\bibfnamefont {P.}~\bibnamefont
  {Gambardella}}\ and\ \bibinfo {author} {\bibfnamefont {I.~M.}\ \bibnamefont
  {Miron}},\ }\bibfield  {title} {\bibinfo {title} {{Current-induced spin-orbit
  torques.}},\ }\href {https://doi.org/10.1098/rsta.2010.0336} {\bibfield
  {journal} {\bibinfo  {journal} {Phil. Trans. R. Soc. A}\ }\textbf {\bibinfo
  {volume} {369}},\ \bibinfo {pages} {3175} (\bibinfo {year}
  {2011})}\BibitemShut {NoStop}%
\bibitem [{\citenamefont {Manchon}\ \emph {et~al.}(2019)\citenamefont
  {Manchon}, \citenamefont {{\v{Z}}elezn{\'{y}}}, \citenamefont {Miron},
  \citenamefont {Jungwirth}, \citenamefont {Sinova}, \citenamefont {Thiaville},
  \citenamefont {Garello},\ and\ \citenamefont {Gambardella}}]{Manchon2019}%
  \BibitemOpen
  \bibfield  {author} {\bibinfo {author} {\bibfnamefont {A.}~\bibnamefont
  {Manchon}}, \bibinfo {author} {\bibfnamefont {J.}~\bibnamefont
  {{\v{Z}}elezn{\'{y}}}}, \bibinfo {author} {\bibfnamefont {I.~M.}\
  \bibnamefont {Miron}}, \bibinfo {author} {\bibfnamefont {T.}~\bibnamefont
  {Jungwirth}}, \bibinfo {author} {\bibfnamefont {J.}~\bibnamefont {Sinova}},
  \bibinfo {author} {\bibfnamefont {A.}~\bibnamefont {Thiaville}}, \bibinfo
  {author} {\bibfnamefont {K.}~\bibnamefont {Garello}},\ and\ \bibinfo {author}
  {\bibfnamefont {P.}~\bibnamefont {Gambardella}},\ }\bibfield  {title}
  {\bibinfo {title} {Current-induced spin-orbit torques in ferromagnetic and
  antiferromagnetic systems},\ }\href
  {https://doi.org/10.1103/RevModPhys.91.035004} {\bibfield  {journal}
  {\bibinfo  {journal} {Rev. Mod. Phys.}\ }\textbf {\bibinfo {volume} {91}},\
  \bibinfo {pages} {035004} (\bibinfo {year} {2019})}\BibitemShut {NoStop}%
\bibitem [{\citenamefont {Haney}\ \emph {et~al.}(2007)\citenamefont {Haney},
  \citenamefont {Waldron}, \citenamefont {Duine}, \citenamefont
  {N{\'{u}}{\~{n}}ez}, \citenamefont {Guo},\ and\ \citenamefont
  {MacDonald}}]{Haney2007}%
  \BibitemOpen
  \bibfield  {author} {\bibinfo {author} {\bibfnamefont {P.~M.}\ \bibnamefont
  {Haney}}, \bibinfo {author} {\bibfnamefont {D.}~\bibnamefont {Waldron}},
  \bibinfo {author} {\bibfnamefont {R.~A.}\ \bibnamefont {Duine}}, \bibinfo
  {author} {\bibfnamefont {A.}~\bibnamefont {N{\'{u}}{\~{n}}ez}}, \bibinfo
  {author} {\bibfnamefont {H.}~\bibnamefont {Guo}},\ and\ \bibinfo {author}
  {\bibfnamefont {a.~H.}\ \bibnamefont {MacDonald}},\ }\bibfield  {title}
  {\bibinfo {title} {{Ab initio giant magnetoresistance and current-induced
  torques in Cr/Au/Cr multilayers}},\ }\href
  {https://doi.org/10.1103/PhysRevB.75.174428} {\bibfield  {journal} {\bibinfo
  {journal} {Phys. Rev. B}\ }\textbf {\bibinfo {volume} {75}},\ \bibinfo
  {pages} {174428(1)} (\bibinfo {year} {2007})}\BibitemShut {NoStop}%
\bibitem [{\citenamefont {Shi}\ \emph {et~al.}(2006)\citenamefont {Shi},
  \citenamefont {Zhang}, \citenamefont {Xiao},\ and\ \citenamefont
  {Niu}}]{Shi2006}%
  \BibitemOpen
  \bibfield  {author} {\bibinfo {author} {\bibfnamefont {J.}~\bibnamefont
  {Shi}}, \bibinfo {author} {\bibfnamefont {P.}~\bibnamefont {Zhang}}, \bibinfo
  {author} {\bibfnamefont {D.}~\bibnamefont {Xiao}},\ and\ \bibinfo {author}
  {\bibfnamefont {Q.}~\bibnamefont {Niu}},\ }\bibfield  {title} {\bibinfo
  {title} {Proper definition of spin current in spin-orbit coupled systems},\
  }\href {https://doi.org/10.1103/PhysRevLett.96.076604} {\bibfield  {journal}
  {\bibinfo  {journal} {Phys. Rev. Lett.}\ }\textbf {\bibinfo {volume} {96}},\
  \bibinfo {pages} {076604} (\bibinfo {year} {2006})}\BibitemShut {NoStop}%
\bibitem [{\citenamefont {Gilbert}(2004)}]{Gilbert2004}%
  \BibitemOpen
  \bibfield  {author} {\bibinfo {author} {\bibfnamefont {T.}~\bibnamefont
  {Gilbert}},\ }\bibfield  {title} {\bibinfo {title} {{Classics in Magnetics A
  Phenomenological Theory of Damping in Ferromagnetic Materials}},\ }\href
  {https://doi.org/10.1109/TMAG.2004.836740} {\bibfield  {journal} {\bibinfo
  {journal} {IEEE Transactions on Magnetics}\ }\textbf {\bibinfo {volume}
  {40}},\ \bibinfo {pages} {3443} (\bibinfo {year} {2004})}\BibitemShut
  {NoStop}%
\bibitem [{\citenamefont {{{\v Z}elezn{\'y}}}()}]{llgcode}%
  \BibitemOpen
  \bibfield  {author} {\bibinfo {author} {\bibfnamefont {J.}~\bibnamefont {{{\v
  Z}elezn{\'y}}}},\ }\href@noop {} {}\bibinfo {howpublished}
  {\url{https://bitbucket.org/zeleznyj/macrospin-llg}}\BibitemShut {NoStop}%
\bibitem [{\citenamefont {Gomonay}\ \emph {et~al.}(2012)\citenamefont
  {Gomonay}, \citenamefont {Kunitsyn},\ and\ \citenamefont
  {Loktev}}]{Gomonay2012}%
  \BibitemOpen
  \bibfield  {author} {\bibinfo {author} {\bibfnamefont {H.~V.}\ \bibnamefont
  {Gomonay}}, \bibinfo {author} {\bibfnamefont {R.~V.}\ \bibnamefont
  {Kunitsyn}},\ and\ \bibinfo {author} {\bibfnamefont {V.~M.}\ \bibnamefont
  {Loktev}},\ }\bibfield  {title} {\bibinfo {title} {{Symmetry and the
  macroscopic dynamics of antiferromagnetic materials in the presence of
  spin-polarized current}},\ }\href
  {https://doi.org/10.1103/PhysRevB.85.134446} {\bibfield  {journal} {\bibinfo
  {journal} {Phys. Rev. B}\ }\textbf {\bibinfo {volume} {85}},\ \bibinfo
  {pages} {134446} (\bibinfo {year} {2012})}\BibitemShut {NoStop}%
\end{thebibliography}%

\end{document}


\renewcommand{\theequation}{S\arabic{equation}}
\renewcommand{\thefigure}{S\arabic{figure}}

\title{Supplemental Material: Robust spin-transfer torque and magnetoresistance in non-collinear antiferromagnetic junctions}
\author{Srikrishna Ghosh}
\affiliation{Institute of Physics, Czech Academy of Sciences, Cukrovarnick\'{a} 10, 162 00 Praha 6 Czech Republic}
\author{Aurelien Manchon}
\affiliation{CINaM, Aix-Marseille Univ, CNRS, Marseille, France}
\author{Jakub \v{Z}elezn\'{y}}
\affiliation{Institute of Physics, Czech Academy of Sciences, Cukrovarnick\'{a} 10, 162 00 Praha 6 Czech Republic}

\maketitle

\tableofcontents

\section{Expression for the torque}

The current induced torque acting on the magnetic moments can be generally expressed using the non-equilibrium (current induced) spin accumulation \cite{Gambardella2011,Manchon2019}. For our system we have
\begin{align}
 \frac{d \mathbf{M}_a}{dt} = \gamma \mathbf{T}_a = -\gamma \mathbf{M}_a \times \frac{J \delta {\boldsymbol \sigma}_a}{M_a} = -\gamma \mathbf{m}_a \times J \delta {\boldsymbol \sigma}_a,\label{eq:torque}
\end{align}
where $\gamma = 2\mu_B/\hbar$ is the electron gyromagnetic ratio, $\mathbf{T}_a$ denotes the torque acting on the magnetic moment $a$ and by $\delta {\boldsymbol \sigma}_a$ we denote the non-equilibrium spin acummulation on site $a$ in the units of $\hbar/2$ (this corresponds to using directly the Pauli matrices for the spin operators). Note that here we consider only the current induced torque and ignore other sources of torque such as the exchange, anisotropy or damping. This expression can be alternatively rewritten as
\begin{align}
 \frac{d \mathbf{M}_a}{dt} = -\frac{2}{ \hbar} \mathbf{m}_a \times J \delta \mathbf{M}_a,\label{eq:torque_cisp}
\end{align}
where $\delta \mathbf{M}_a$ = $\mu_B\delta {\boldsymbol \sigma}$ is the current induced magnetization on site $a$.

From Eq. \eqref{eq:torque} we see that the torque can be either evaluated using the non-equilibrium spin accumulation or using a torque operator
\begin{align}
  \hat{\mathbf{T}}_a = -\mathbf{m_a} \times J \hat{\boldsymbol \sigma}_a,
 \end{align}
where $\hat{\boldsymbol \sigma}_a$ is the (dimensionless) spin-operator projected on site $a$.

Alternatively, in the non-relativistic limit the torque can be expressed using the spin current divergence (spin sources) \cite{Haney2007,Manchon2019}. To see how this relates to the expression based on the current induced spin accumulation we consider that in general it holds for a wavefunction $\psi$ \cite{Shi2006}
\begin{align}
 \frac{\partial \bra{\psi(\mathbf{r})}\hat{\sigma}_i\ket{\psi(\mathbf{r})}}{\partial t}  + {\boldsymbol \nabla}\cdot \bra{\psi(\mathbf{r})}\hat{\bm{\mathcal J}}^i\ket{\psi(\mathbf{r})} = \frac{1}{i\hbar}\bra{\psi(\mathbf{r})}[\sigma_i,\hat{H}]\ket{\psi(\mathbf{r})}\label{eq:spin_continuity}
\end{align}
Here $\hat{\bm{\mathcal J}}^i = \frac{1}{2}\{\hat{\sigma}_i,\hat{\mathbf{v}}\}$ is the spin current operator ($\hat{\mathbf{v}}$ denotes the velocity operator). For our system the commutator on the right hand side can be expressed as
\begin{align}
 \frac{1}{i\hbar}[\hat{\boldsymbol \sigma},\hat{H}] = \frac{2}{\hbar} \sum_a \mathbf{m_a} \times J \hat{\boldsymbol \sigma}_a = -\frac{2}{\hbar} \hat{\mathbf{T}},
 \label{eq:spin_commutator}
\end{align}
where $\hat{\mathbf{T}}$ is the total torque operator. In a steady state the first term in Eq. \eqref{eq:spin_continuity} must vanish.
Thus we see that source of spin current directly correspond to the torque:
\begin{align}
 {\boldsymbol \nabla} \bm{\mathcal J}^i(\mathbf{r}) = -\frac{2}{\hbar} \mathbf{T}(\mathbf{r})
\end{align}
where $\bm{\mathcal J}^i$ and $\mathbf{T}$ correspond to the mean values of spin current and torque respectively.

For the tight-binding model that we use the correspondence has a simple form. The change of spin on site $a$ for state $\psi$ is given by:
\begin{align}
\frac{d \bra{\psi}\hat{\boldsymbol \sigma}_a \ket{\psi } } { dt} = \bra{\psi} \frac{1}{i\hbar} [\hat{\boldsymbol \sigma}_a,\hat{H}]\ket{\psi}.
\label{eq:spin_evolution}
\end{align}

In steady state the left-hand side has to vanish. In block form, the Hamiltonian can be expressed as:
\begin{align}
 \hat{H} = 
    \left(\begin{matrix}
     H_{aa} & H_{ab} & \dots \\
     H_{ba} & H_{bb} & \dots \\
     \dots & \dots & \dots
    \end{matrix}\right),
\end{align}
where $H_{aa}$ denotes the Hamiltonian submatrix corresponding to matrix elements on site $a$ and analogously for the others. $\hat{\boldsymbol \sigma}_a$ can be similarly written as:

\begin{align}
  \hat{\boldsymbol \sigma}_a = 
    \left(\begin{matrix}
     {\boldsymbol \sigma} & 0 & \dots \\
     0 & 0 & \dots \\
     \dots & \dots & \dots
    \end{matrix}\right).
\end{align}
Then the commutator can be expressed as 
\begin{align}
 [\hat{\boldsymbol \sigma}_a,\hat{H}] = 
    \left(\begin{matrix}
        [{\boldsymbol \sigma},H_{aa}] & {\boldsymbol \sigma} H_{ab} & \dots \\
        -H_{ba} {\boldsymbol \sigma}   & 0 & \dots \\
        \dots & \dots & \dots
    \end{matrix}\right).
\end{align}
From Eq. \eqref{eq:spin_commutator} we have 
\begin{align}
 [{\boldsymbol \sigma},H_{aa}] = \frac{2}{i} \hat{\mathbf{T}}_a
\end{align}
and using the fact that Eq. \eqref{eq:spin_evolution} has to vanish we have:
\begin{align}
 2 \bra{\psi} \hat{\mathbf{T}}_a \ket{\psi} = i \sum_b \psi_b^*H_{ba}{\boldsymbol \sigma}\psi_a - \psi_a^* {\boldsymbol \sigma}H_{ab}\psi_b.
 \label{eq:torque_bond_currents}
\end{align}

The expression corresponds to the so-called (spin) bond-currents. They are described by the operator:
\begin{align}
\hat{I}_{ab}^i = 
    i\left(\begin{matrix}
        0 & -\sigma^i H_{ab} & \dots \\
        H_{ba} \sigma^i   & 0 & \dots \\
        \dots & \dots & \dots
    \end{matrix}\right).
\end{align}
The bond currents directly give the spin current operator:
\begin{align}
 \hat{\bm{\mathcal J}}_a^i = \frac{1}{2\hbar} \sum_b (\mathbf{x}_j - \mathbf{x}_i) \hat{I}_{ab}^i 
\end{align}
From Eq. \eqref{eq:torque_bond_currents} we have:
\begin{align}
 \hat{\mathbf{T}}_a = \frac{1}{2} \sum_b \hat{I}_{ab}.
\end{align}
The expression on the right-hand side can be understood as spin source, i.e. it gives the amount of spin carried by the spin current out of the atom $a$ per unit of time.

\section{Switching simulations}

We simulate the dynamics induced by the spin transfer torque using the semiclassical Landau-Lifschitz-Gilbert (LLG) equations \cite{Gilbert2004} within the macrospin approximation, i.e. we assume that the magnetic order is homogeneous and no magnetic domains are present. The code we used for the simulations is available at \cite{llgcode}. To describe the magnetic interactions in the system we consider a nearest neighbor antiferromagnetic exchange, which we set to $100\ \text{meV}$ and uniaxial anisotropy on each site set to $0.1\ \text{meV}$. The directions of the uniaxial anisotropy on each site are rotated with $120^\circ$ to each other and are set so that the moments as shown in Fig. 1 in the main text are oriented along the anisotropy axes. We also include damping with Gilbert damping parameter $\alpha=0.01$. We set the magnitude of the magnetic moment to $M=3\mu_\text{B}$. Since we consider a large exchange (corresponding to high N\'eel temperature), the magnetic order will always stay close to the triangular magnetic order during the dynamics.

We simulate the dynamics for the in-plane and out-of-plane torque separately since these torques have a distinct behavior both in regards to the induced dynamics and regarding their microscopic origin. We assume that the left layer is fixed and simulate the dynamics in the right magnetic layer. For the right layer we consider as initial states either the configuration shown in Fig. 1(a) of the main text or the reversed state, which correspond to parallel and anti-parallel junctions, respectively. We first discuss the in-plane torque. The in-plane torque drives the magnetic order out of the plane. This is because the in-plane torque causes an in-plane canting of the magnetic moments, which then results in a large exchange torque, which drives the magnetic moment out of the plane. Since our calculations are only for the in-plane orientation of the magnetic moments, we cannot use the calculations directly for the simulations. Instead we use the fact that the in-plane torque is well described by the $\mathbf{T}^\text{loc}$ and $\mathbf{T}^\text{AD}$ torques as described in the main text. This thus gives as description of the torque for any orientation of the moments. Note that when the magnetic moments are tilted out of the plane, the $\mathbf{T}^\text{loc}$ and $\mathbf{T}^\text{AD}$ will no longer be in-plane.

\begin{figure}
\includegraphics[width=\columnwidth]{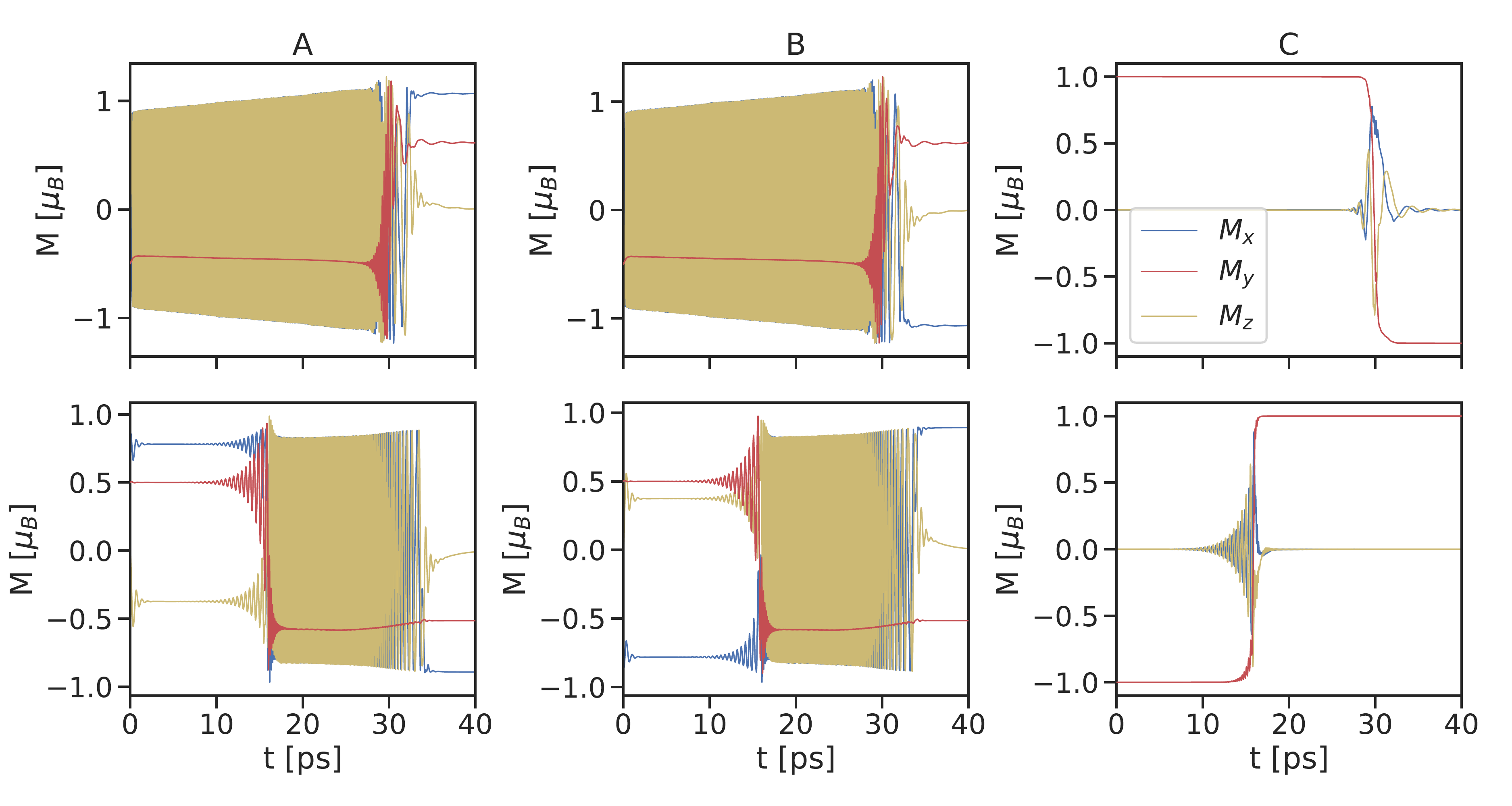}%
\caption{\label{fig:dynamics_ip} \textbf{(a)} Simulations of switching by the $\mathbf{T}^\text{loc}$ and $\mathbf{T}^\text{AD}$ torques. Here the top panels show the switching of the right layer from the parallel configuration to the antiparallel configuration and vice versa for the bottom panels. The sign of the bias and thus also of the torque is reversed between the top and bottom panels. Here  $|\mathbf{T}^\text{AD}| = 0.04\ \text{eV}$ and $|\mathbf{T}^\text{loc}| = 0.07\ \text{eV}$, which corresponds to the calculated torque for  $E_F = -1.25\ \text{eV}$, $D=0.2\ \text{eV}$ and $n_\text{steps} = 25$ and 0.01 eV bias. In both cases we find rapid oscillations in the $x-z$ plane, which correspond to the yellow regions.}
\end{figure}

\begin{figure}
\includegraphics[width=\columnwidth]{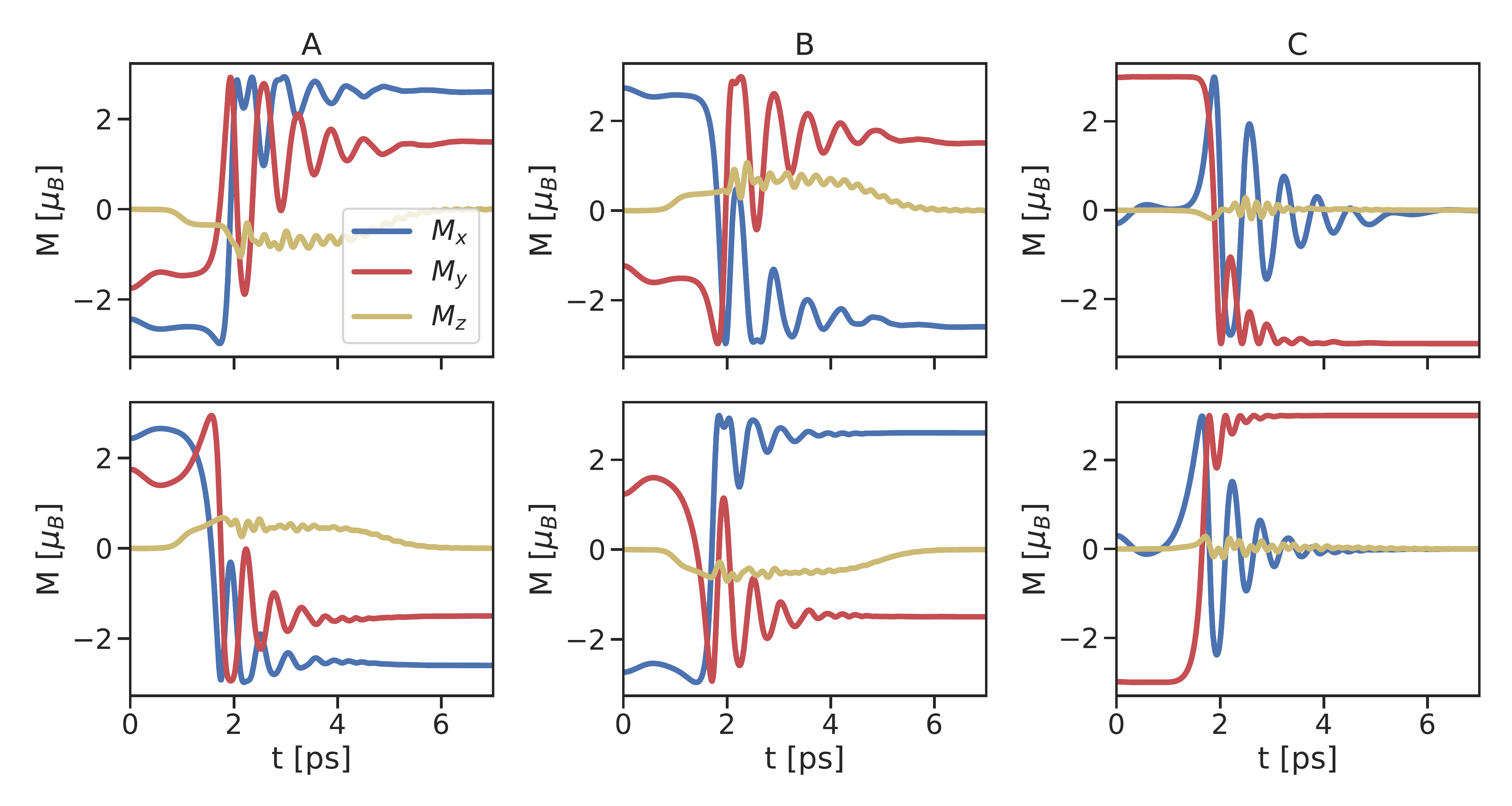}%
\caption{\label{fig:dynamics} Simulations of switching by the calculated torque with easy-plane anisotropy. Here the top panels show the switching of the right layer from the parallel configuration to the antiparallel configuration and vice versa for the bottom panels. The sign of the bias and thus also of the torque is reversed between the top and bottom panels. Here  $|\mathbf{T}^\text{AD}| = 0.04\ \text{eV}$ and $|\mathbf{T}^\text{loc}| = 0.07\ \text{eV}$, which corresponds to the calculated torque for  $E_F = 1.5\ \text{eV}$, $D=0.2\ \text{eV}$ and $n_\text{steps} = 25$ and 0.01 eV bias.}
\end{figure} 

In Fig. \ref{fig:dynamics_ip} we show a simulation of switching between the parallel and antiparallel states by the $\mathbf{T}^\text{loc}$ and $\mathbf{T}^\text{AD}$ torques corresponding to the case of $E_F = -1.25\ \text{eV}$, $D=0.2\ \text{eV}$ and $n_\text{steps} = 25$ as shown in Fig. 2 of the main text.  The torque is gradually switched off around $t=35\ \text{ps}$.  These simulations illustrate that the torque in our system can switch the junction between the parallel and antiparallel states using current pulses with opposite directions. The exact timing of the pulse is not important here, which demonstrates that the switching is deterministic. The switching here is caused mainly by the $\mathbf{T}^\text{AD}$ torque. We find that even if the $\mathbf{T}^\text{loc}$ is set to zero, switching is still possible. However, we find that the  $\mathbf{T}^\text{loc}$ can allow for switching with smaller $\mathbf{T}^\text{AD}$ than what would be otherwise possible. The relationship between the $\mathbf{T}^\text{loc}$ and $\mathbf{T}^\text{AD}$ torques appears to be complicated and exploring it in detail is left to future work. Note that the simulation starts here from the exact parallel or anti-parallel states, illustrating that no thermal fluctuation is needed to initiate the switching, unlike in the FM case. The critical switching torque is in this case sensitive both by the anisotropy and by the Gilbert damping, as expected for anti-damping torque \cite{Gomonay2012}. The exact determination of the relation between the anisotropy, damping and the critical torque is left for future work.

To simulate switching due to the $T_z$ torque we  consider an additional easy-plane anisotropy that constrains the moments to the plane, which we set to $5\ \text{meV}$. In such a case the magnetic moments always stay close to the calculated state and then we can directly use the calculated torque for the simulations. We consider the average torque per site in the right magnetic layer for each magnetic sublattice and $E_F = 1.5\ \text{eV}$, $D=0.2\ \text{eV}$ and $n_\text{steps} = 25$ as shown in Fig. \ref{fig:sublattice_fit_1}. We consider all the torque components here, but the in-plane components are rendered ineffective by the easy-plane anisotropy.

In Fig. \ref{fig:dynamics} we show the switching between the parallel and anti-parallel states. We see that, analogously to the in-plane torque, the $T_z$ torque can switch the junction between the parallel and antiparallel states using current pulses with opposite directions. We note that for the dynamics we gradually switch on the torque so that it becomes maximal around $t=2\ \text{ps}$ and also gradually switch it off so that the torque vanishes around $t=6\ \text{ps}$. Exact timing of the pulses is not important here for the switching. Since the torque would vanish for the parallel or anti-parallel configuration we consider a small rotation of the moments for the initial states. This is similar to spin-transfer torque switching of ferromagnets, where a thermal fluctuation away from the easy axis is typically required to initialize the switching.

These simulations include all the calculated torque components, however the switching is caused only by the $T_z$ component of the torque. The dynamics is almost unchanged when the in-plane components of the torque are ignored. This is because the in-plane component drives the magnetic-order out of the plane, however, this is prevented by the out-of-plane anisotropy. The fact that the in-plane torque drives the moments out of the plane and the out-of-plane torque drives them in-plane is a consequence of the strong antiferromagnetic exchange. The STT will initially cause a small tilt of the individual moments with respect to each other, which then results in an exchange torque, which drives the moments in a different direction than the STT. 

The critical switching torque magnitude below which no switching occurs corresponds in this case only to the torque due to magnetocrystalline anisotropy. This is expected since the $T_z$ component of the torque is essentially a field-like torque (at least the total) for which it is known in ferromagnets and collinear antiferromagnets that for the switching the torque has to overcome the anisotropy.

\section{Interfacial disorder}

To incorporate the interfacial disorder we use a random walk along the interface. We describe the interface with a function $I(y)$ which gives the x-position of the interface as a function of the $y$ coordinate. We discretize the $y$ range using a large number of points. Then, starting at $y=0$, we randomly choose at each point whether to stay at the same $x$ position or whether to include an interfacial step in the + or - directions. We use steps with the width of $1/4$ of unit cell since this corresponds to an atomic step. To prevent large deviations from the mean value of $x$ we bias the random walk so that the farther away we are from the mean, the less likely it is to have a step away from the mean.

The probability of having a step at each point is $p = n_\text{steps}/N$, where $N$ is the number of points we use for discretizing the $y$ coordinate. The decision at a point is done by using a random number $r$ from range [0,1] and a deviation $d$, which denotes the deviation of the previous point from the mean value divided by the step size:
\begin{itemize}
 \item If $r < \left(\frac{d}{|d| + 2} + 1\right)\frac{p}{2}$ make a step in the $-x$ direction.
 \item If $r > 1 - \left(p - \left(\frac{d}{|d| + 2} + 1\right)\frac{p}{2}\right)$ make a step in the $+x$ direction.
 \item In other cases, we keep the same $x$ value as in the previous step.
\end{itemize}

\begin{figure}
\includegraphics[width=\columnwidth]{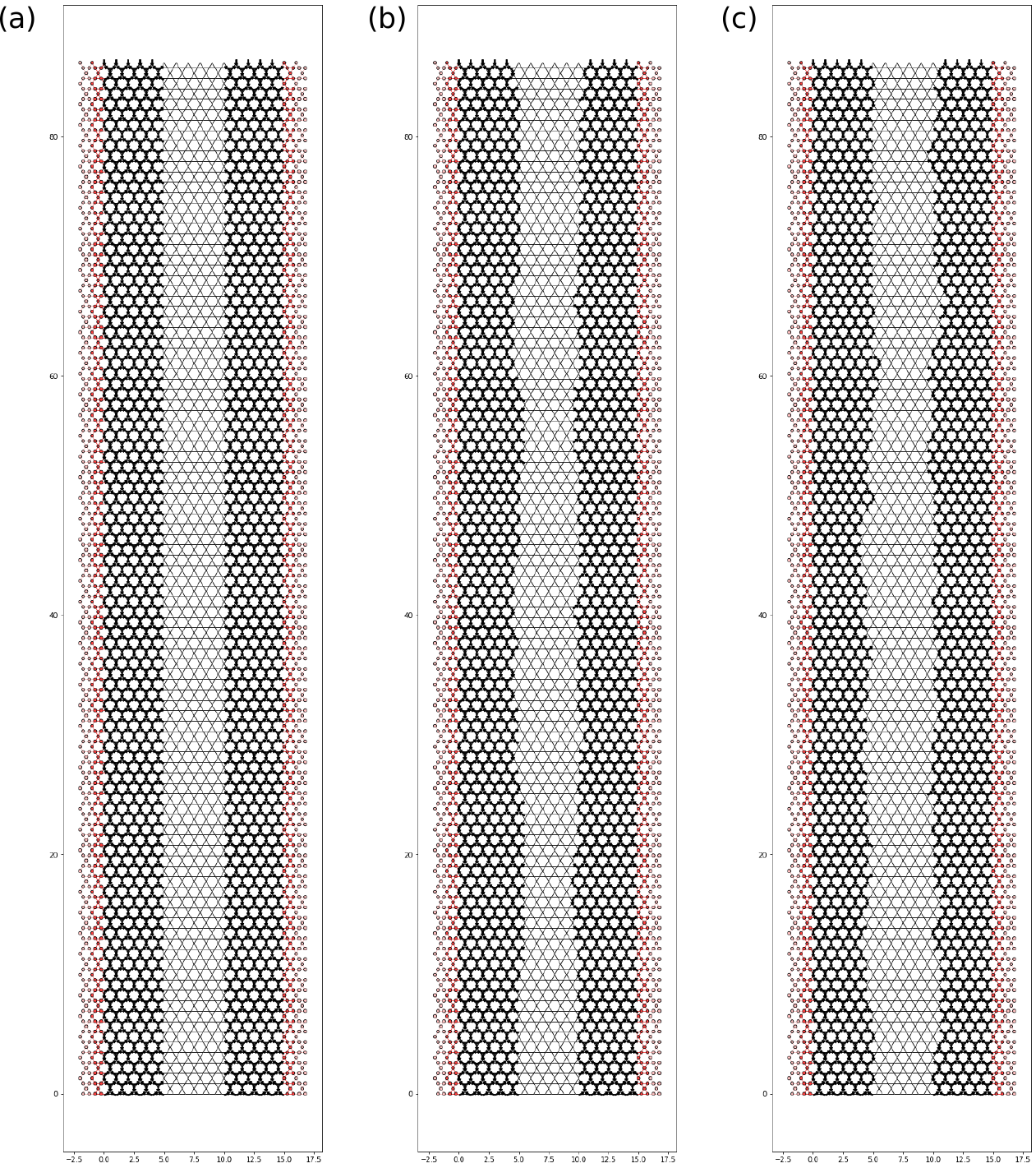}%
\caption{\label{fig:int_disorder}  Illustration of the full size of the junction with the interfacial disorder included. \textbf{(a)} No interfacial disorder. \textbf{(b)} $n_\text{steps} = 25$ \textbf{(c)} $n_\text{steps} = 100$.} 
\end{figure}

In Fig. \ref{fig:int_disorder} we show an illustration of the full size of the junction with $n_\text{steps}$ = 0, $n_\text{steps}$ = 25 and $n_\text{steps}$ = 100.

\section{Additional calculations}

\subsection{$x$ depedence of the spin current}

\begin{figure}
\includegraphics[width=\columnwidth]{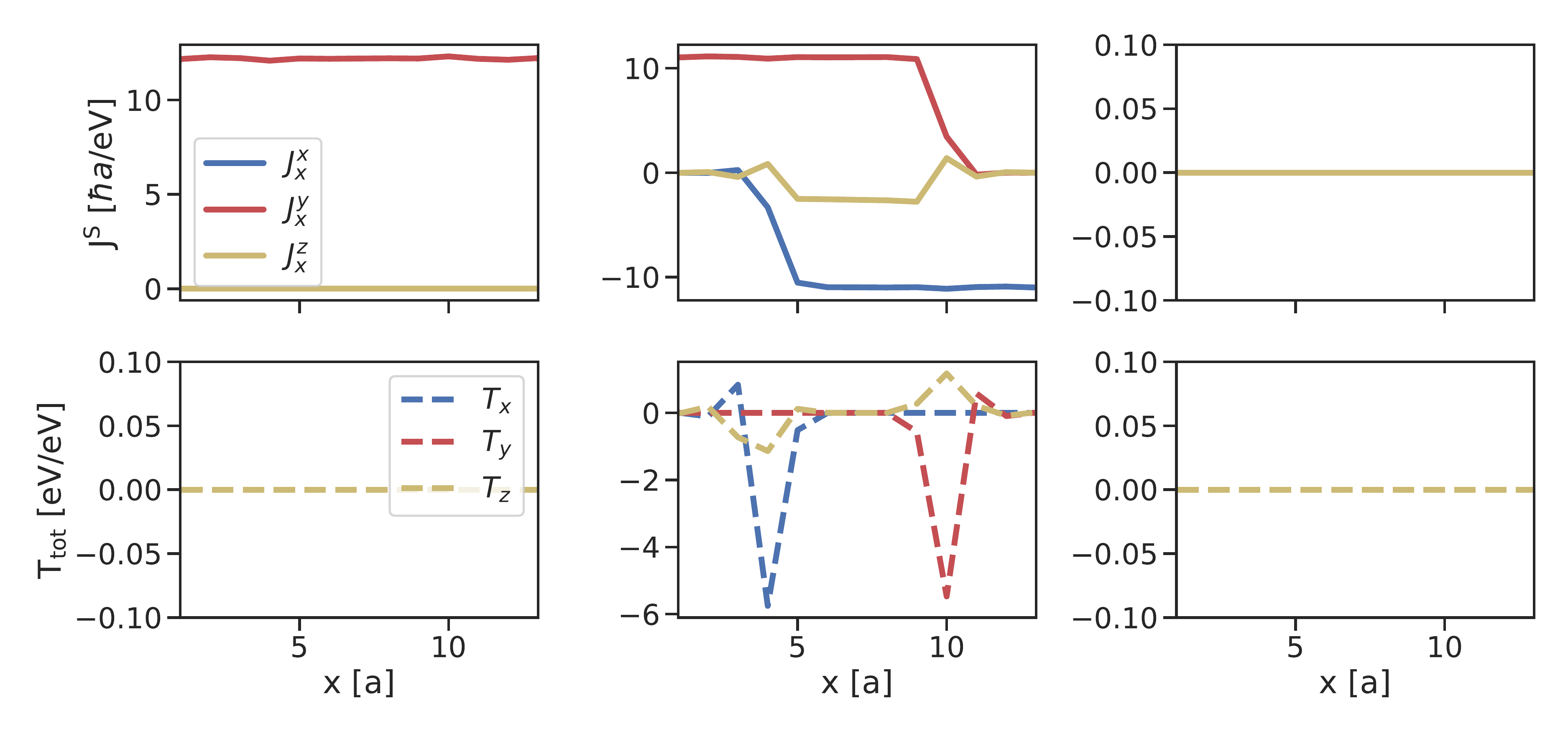}%
\caption{\label{fig:x_profile_FM}  The spin current and torque for the parallel (left), perpendicular (middle), and anti-parallel (right) configurations of the FM junction as a function of the $x$ coordinate ($a$ denotes the lattice constant along the $x$ direction). We sum up both the torque and the spin current within each unit cell and then sum up also along the $y$ direction. $J_x^i$ denote spin current flowing along the $x$ direction with spin-polarization along $i$. We set $D = 0.2\ \text{eV}$, $n_\text{steps}=25$ and $E_F=1.5\ \text{eV}$.}
\end{figure}

\begin{figure}
\includegraphics[width=\columnwidth]{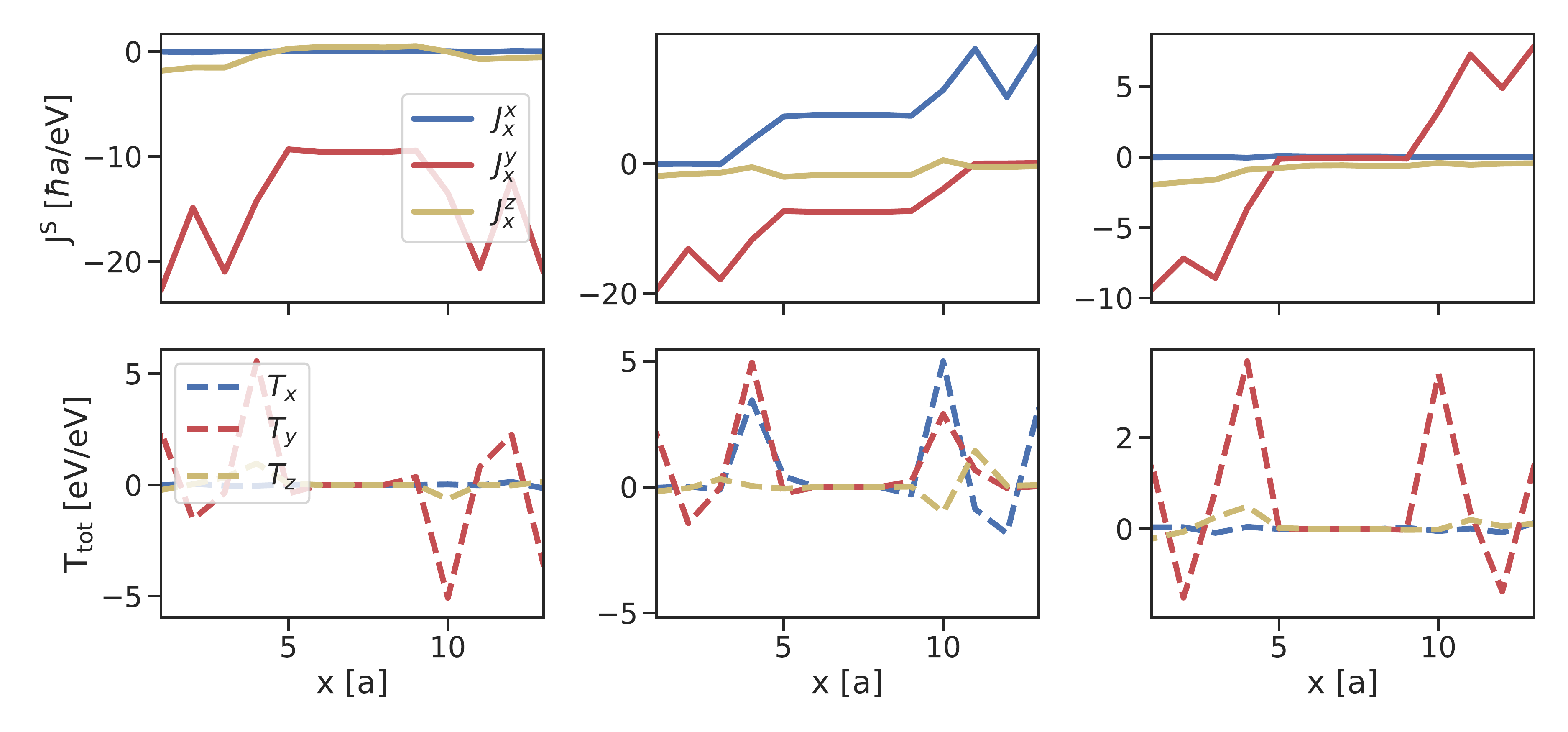}%
\caption{\label{fig:x_profile_nmleads} The spin current and torque for the parallel (left), perpendicular (middle), and anti-parallel (right) configurations of the non-collinear AFM junction with non-magnetic leads as a function of the $x$ coordinate ($a$ denotes the lattice constant along the $x$ direction). We sum up both the torque and the spin current within each unit cell and then sum up also along the $y$ direction. $J_x^i$ denote spin current flowing along the $x$ direction with spin-polarization along $i$. We set $D = 0.2\ \text{eV}$, $n_\text{steps}=25$ and $E_F=1.5\ \text{eV}$.}
\end{figure}

In Fig. \ref{fig:x_profile_FM} we give the $x$-dependence of the spin current and the STT for the FM junction. In this case the torque is present only for the perpendicular configuration of the junction. In Fig. \ref{fig:x_profile_nmleads} we give the $x$-dependence of the spin current and the STT for the non-collinear AFM junction in the case of non-magnetic leads. We see that the case of non-magnetic leads is quite different, but again the STT is directly connected to the absorption of the spin-polarized current.

\subsection{GMR}

\begin{figure}
\includegraphics[width=\columnwidth]{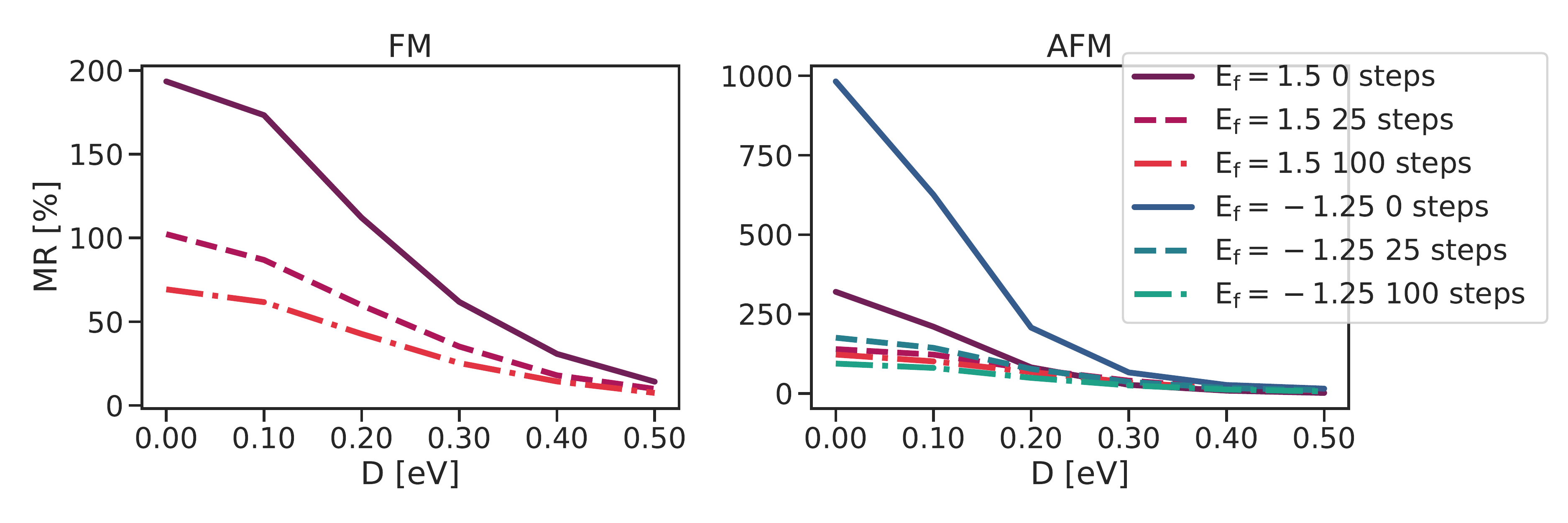}%
\caption{\label{fig:MR_magnitude_nmleads} The dependence of GMR on disorder for the FM and AFM junctions with non-magnetic leads. }
\end{figure}

In Fig. \ref{fig:MR_magnitude_nmleads} we give the magnetoresistance for the case of non-magnetic leads.

\subsection{STT magnitude}

\begin{figure}
\includegraphics[width=\columnwidth]{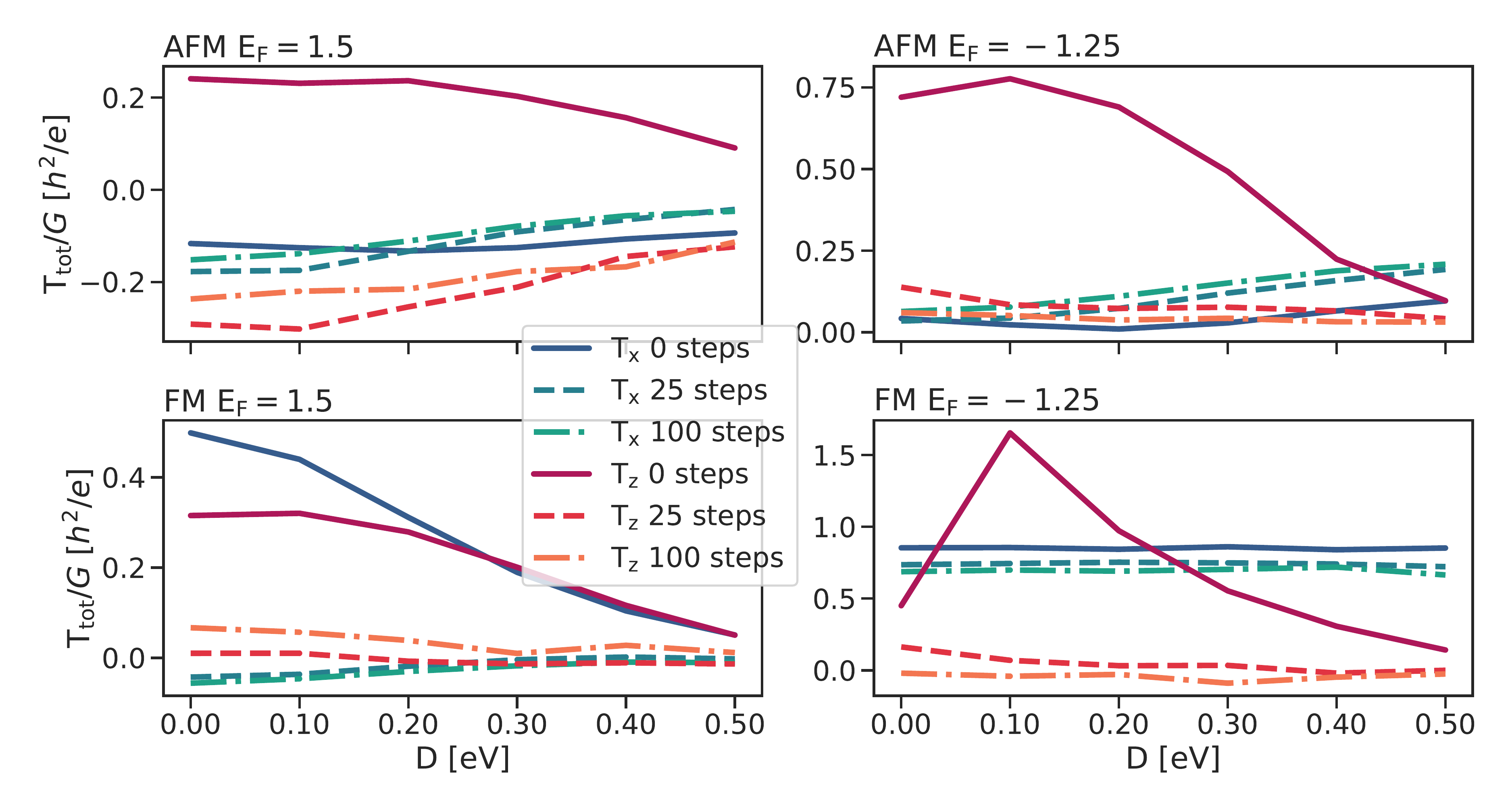}%
\caption{\label{fig:torque_magnitude_nmleads}  The dependence of the torque magnitude scaled by the conductance on the disorder paramater $D$ and on the interfacial disorder characterized by number of interfacial steps for the FM and AFM junction with non-magnetic leads and for two values of $E_F$. The $T_y$ component is not shown here since it has a similar dependence on magnitude as $T_x$. For $T_z$ we set $\theta=90^\circ$; for $T_x$ $\theta=135^\circ$ for the  FM and $\theta=90^\circ$ for the AFM.  }
\end{figure}

In Fig. \ref{fig:torque_magnitude_nmleads} we give the overall magnitude of the STT as a function of disorder for the case of non-magnetic leads.

\newpage
\subsection{Angular dependence of the STT}

\begin{figure}
\includegraphics[width=\columnwidth]{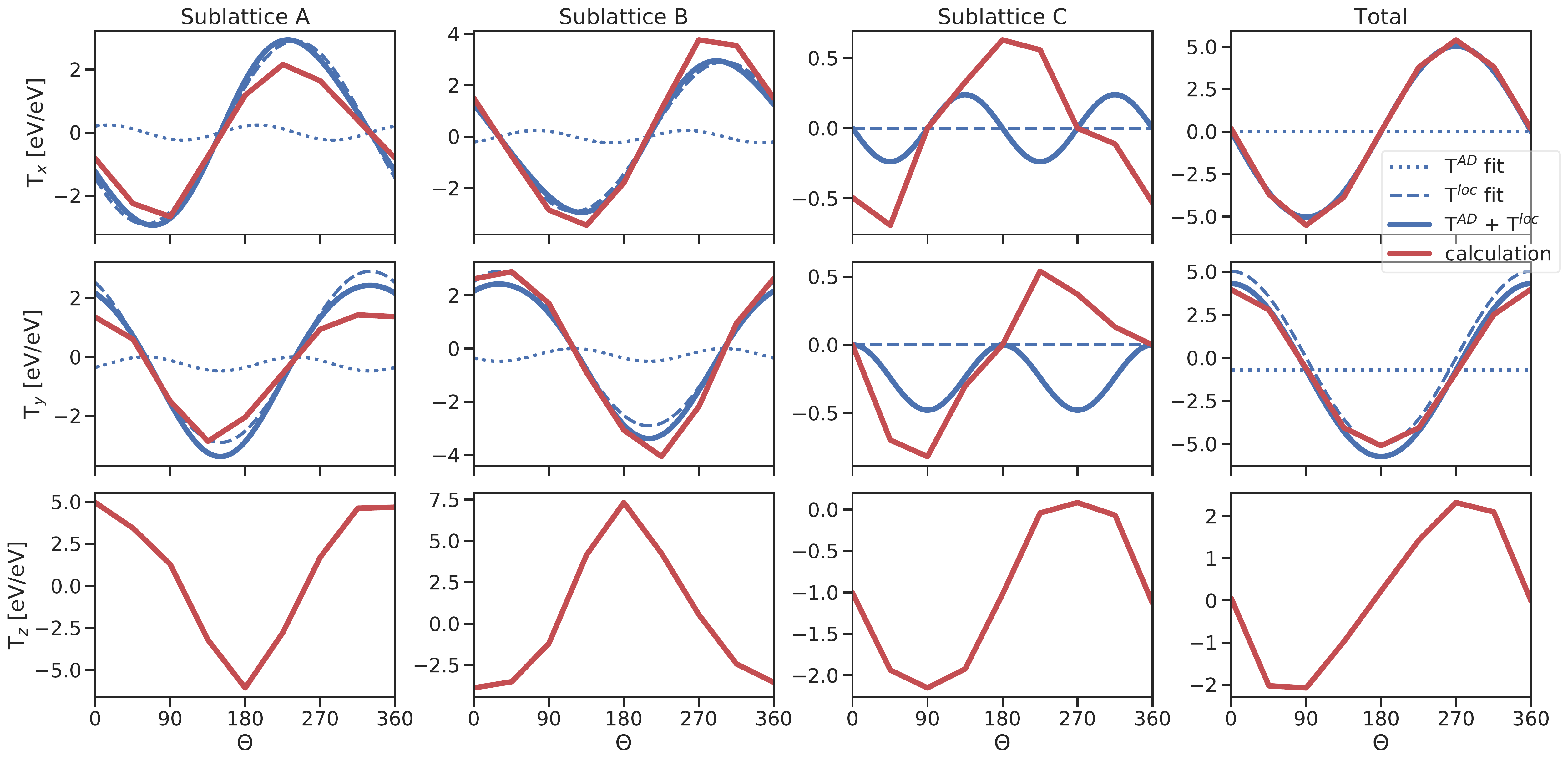}%
\caption{\label{fig:sublattice_fit_1} The dependence of sublattice and the total torque in the right magnetic layer as a function of the rotation angle of the right magnetic layer. The in-plane components of the torque are fitted by the combination of $\mathbf{T}^\text{loc}$ and $\mathbf{T}^\text{AD}$. Here $E_F = 1.5\ \text{eV}$, $D=0.2\ \text{eV}$ and $n_\text{steps} = 25$. }
\end{figure}

\begin{figure}
\includegraphics[width=\columnwidth]{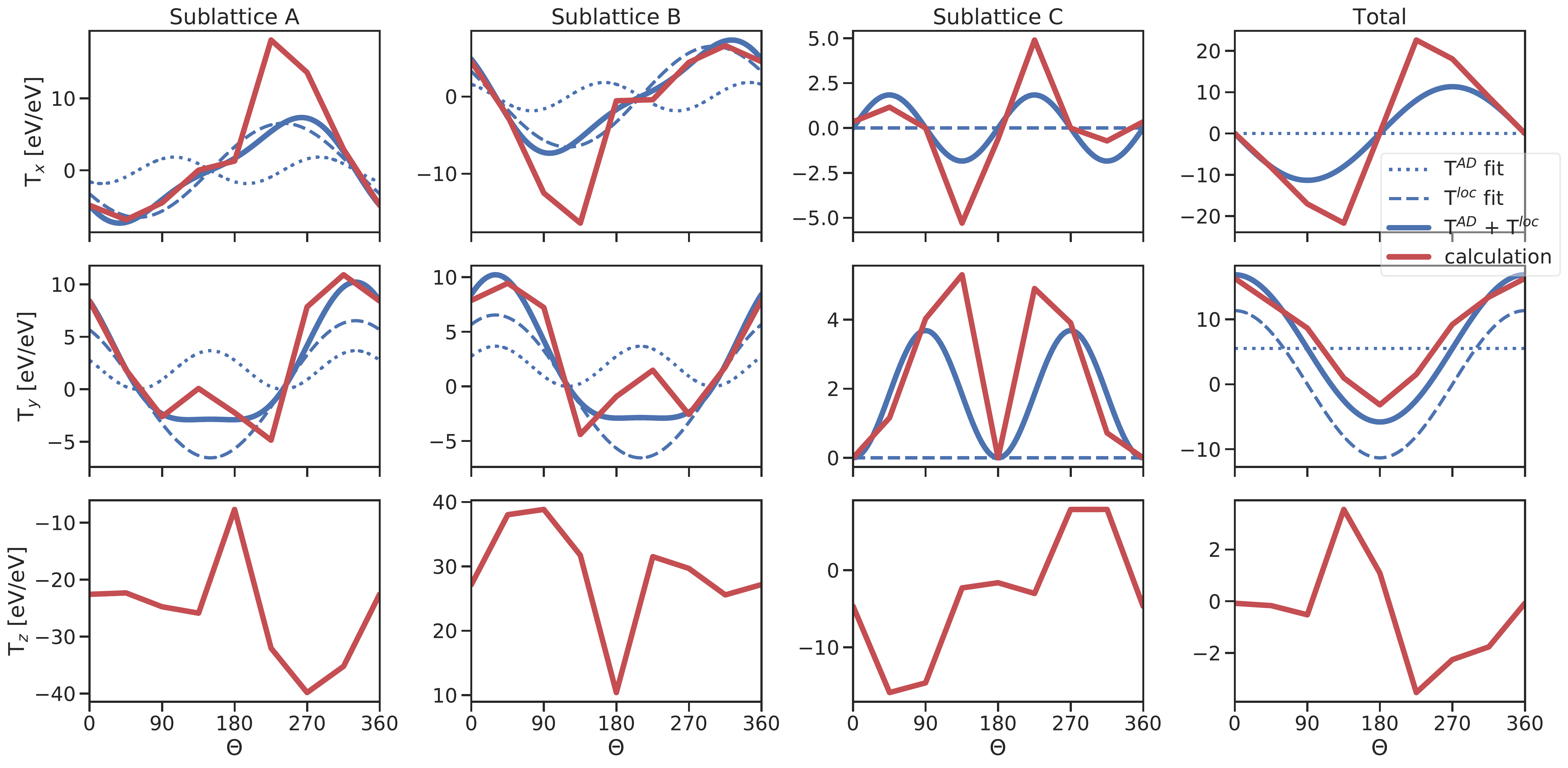}%
\caption{\label{fig:sublattice_fit_2} The dependence of sublattice and the total torque in the right magnetic layer as a function of the rotation angle of the right magnetic layer. The in-plane components of the torque are fitted by the combination of $\mathbf{T}^\text{loc}$ and $\mathbf{T}^\text{AD}$. Here $E_F = -1.25\ \text{eV}$, $D=0.0\ \text{eV}$ and $n_\text{steps} = 0$. }
\end{figure}

\begin{figure}
\includegraphics[width=\columnwidth]{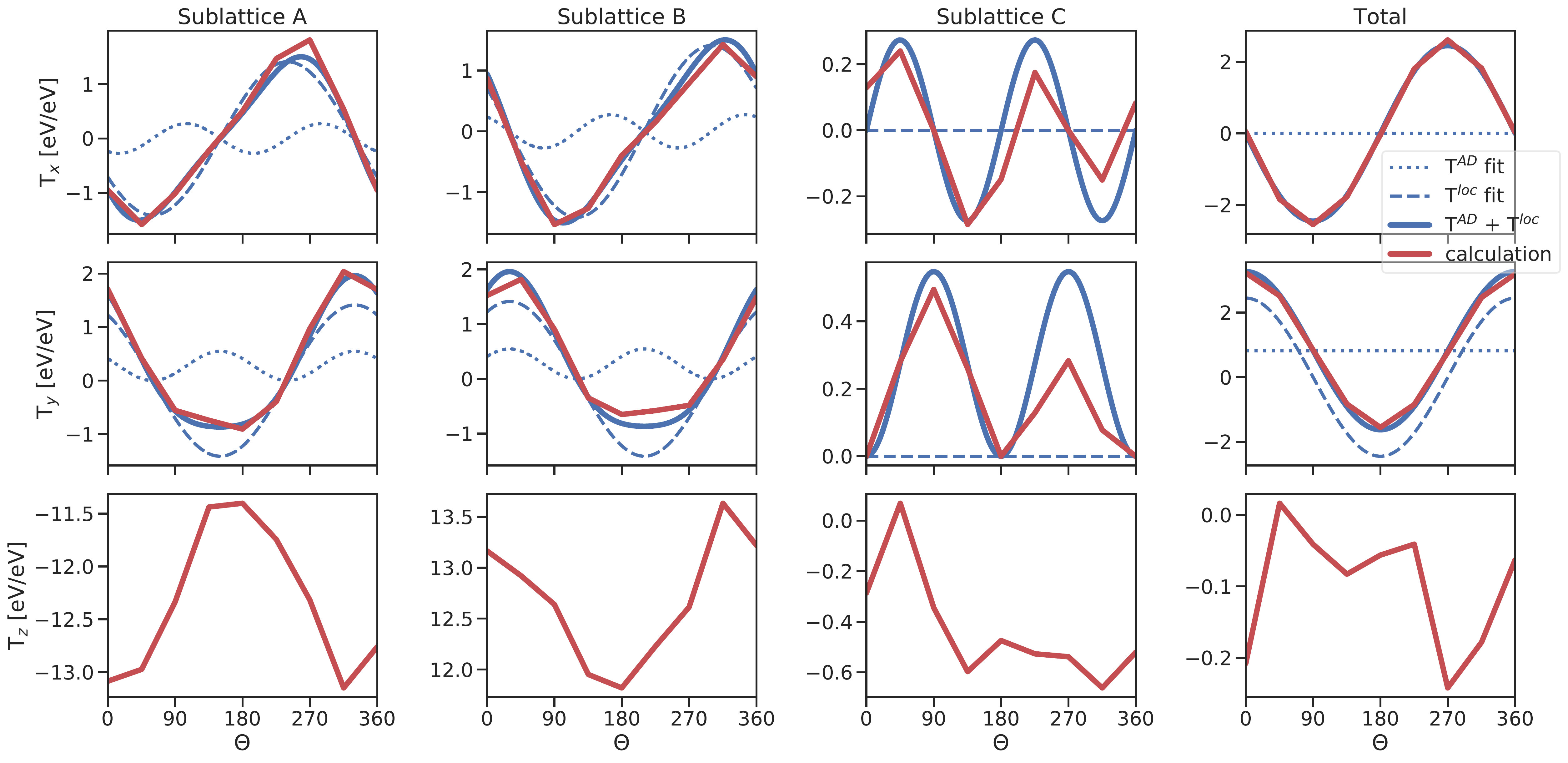}%
\caption{\label{fig:sublattice_fit_3} The dependence of sublattice and the total torque in the right magnetic layer as a function of the rotation angle of the right magnetic layer. The in-plane components of the torque are fitted by the combination of $\mathbf{T}^\text{loc}$ and $\mathbf{T}^\text{AD}$. Here $E_F = -1.25\ \text{eV}$, $D=0.5\ \text{eV}$ and $n_\text{steps} = 100$. }
\end{figure}

\begin{figure}
\includegraphics[width=\columnwidth]{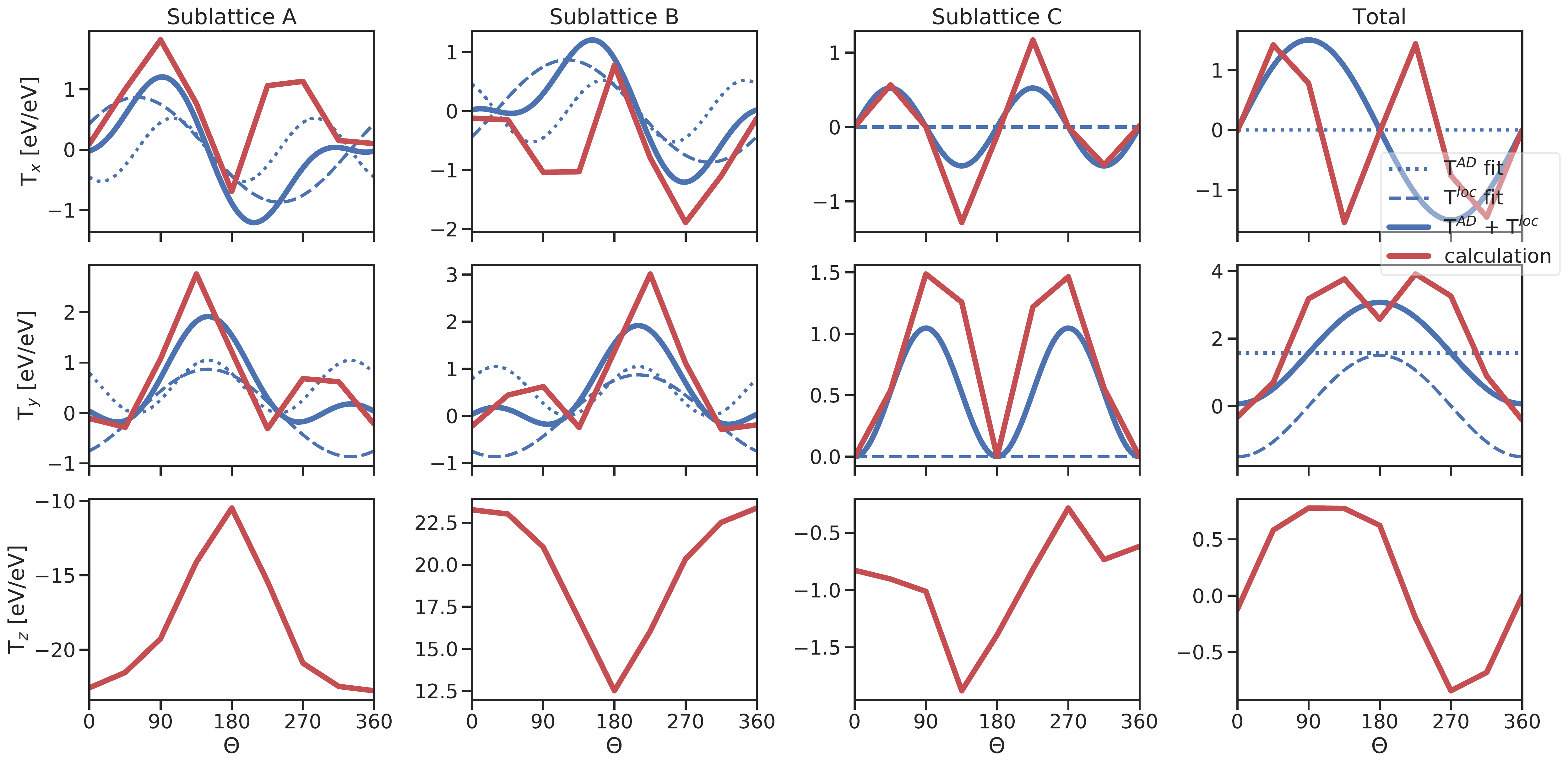}%
\caption{\label{fig:sublattice_fit_4} The dependence of sublattice and the total torque in the right magnetic layer as a function of the rotation angle of the right magnetic layer for a system with non-magnetic leads. The in-plane components of the torque are fitted by the combination of $\mathbf{T}^\text{loc}$ and $\mathbf{T}^\text{AD}$. Here $E_F = -1.25\ \text{eV}$, $D=0.2\ \text{eV}$ and $n_\text{steps} = 25 $. }
\end{figure}

\begin{figure}
\includegraphics[width=\columnwidth]{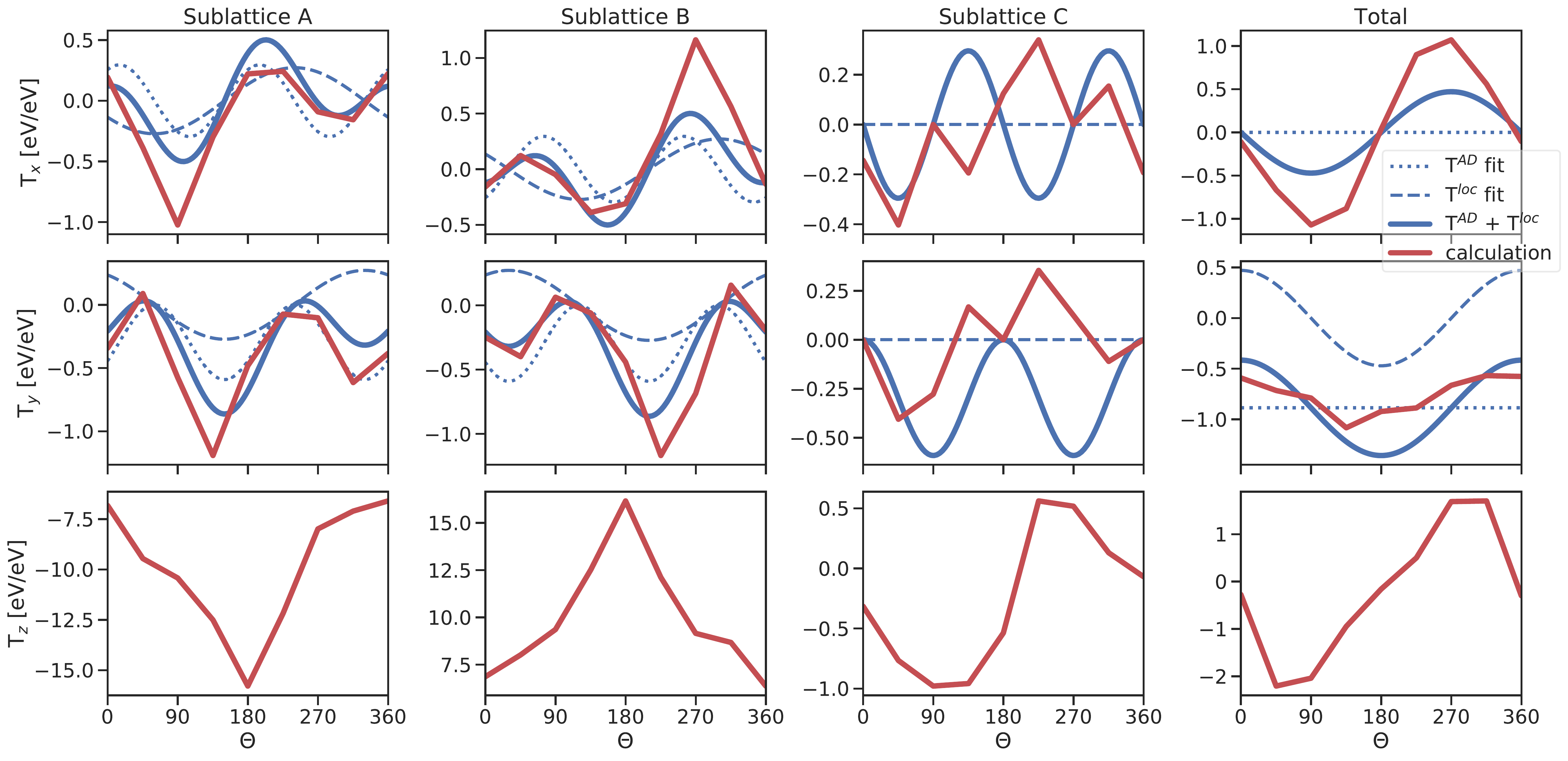}%
\caption{\label{fig:sublattice_fit_5} The dependence of sublattice and the total torque in the right magnetic layer as a function of the rotation angle of the right magnetic layer for a system with non-magnetic leads. The in-plane components of the torque are fitted by the combination of $\mathbf{T}^\text{loc}$ and $\mathbf{T}^\text{AD}$. Here $E_F = 1.5\ \text{eV}$, $D=0.2\ \text{eV}$ and $n_\text{steps} = 25 $. }
\end{figure}

In Figs. \ref{fig:sublattice_fit_1}-\ref{fig:sublattice_fit_5} we give the angular dependence of the torque in the right layer for the antiferromagnetic junctions, in analogy to Fig. 2 of the main text, for various combinations of $E_F$, $D$, $n_\text{steps}$ and for non-magnetic leads. These calculations show that for the case of magnetic leads the torque on sublattices $A$ and $B$ is well fitted by the combination of $\mathbf{T}^\text{loc}$ and $\mathbf{T}^\text{AD}$. The fit is worse on sublattice $C$, however, the torque is typically much smaller here and thus the overall agreement is good. For  the non-magnetic the fit is not as good, although even here, it gives at least a rough description of the torque. 

\begin{figure}
\includegraphics[width=\columnwidth]{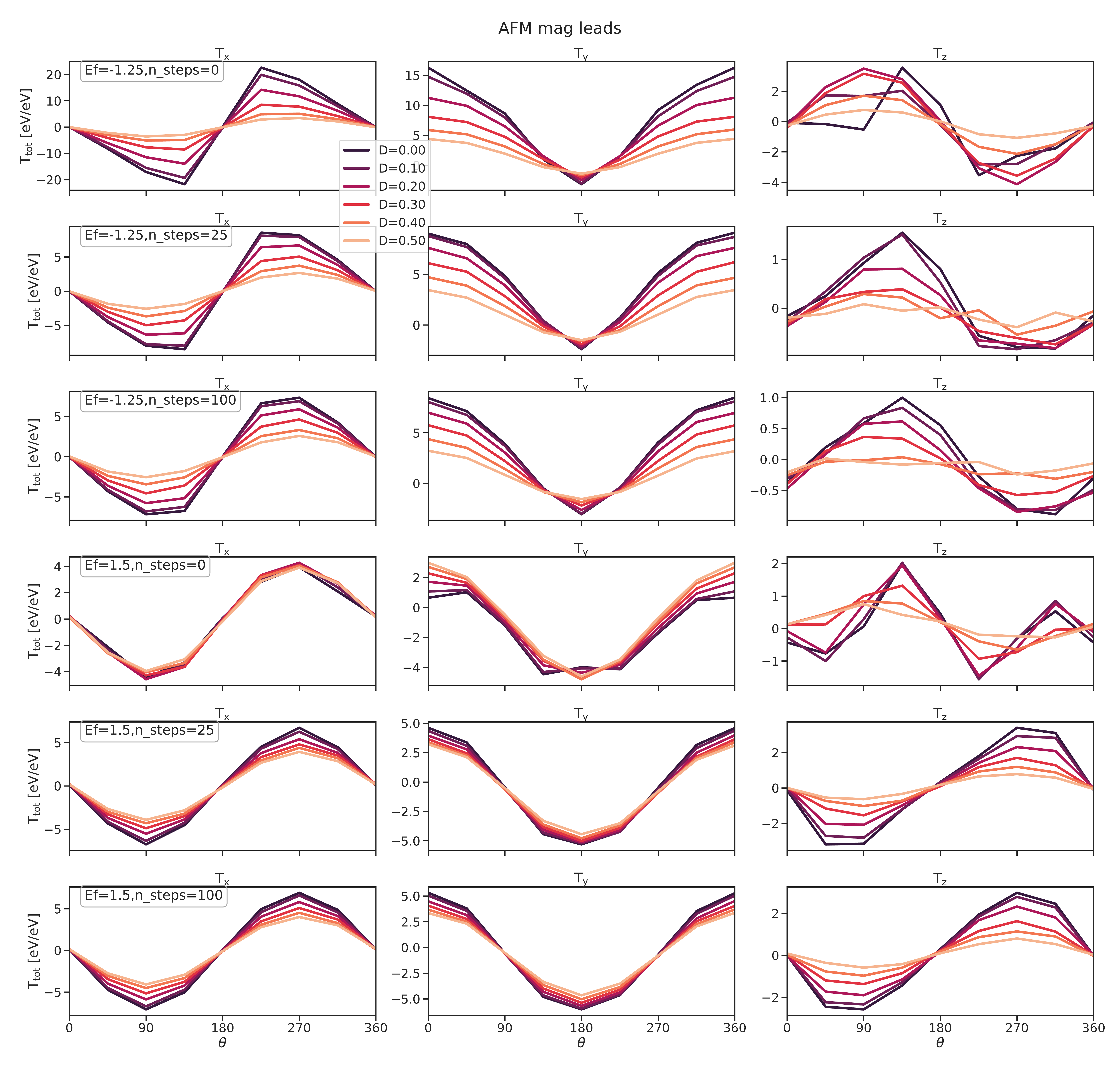}%
\caption{\label{fig:theta_dependence_NCAFM_mag} The dependence of the total torque in the right magnetic region on the relative orientation of the two magnetic layers in the non-collinear AFM junction with magnetic leads for various values of $E_F$ and $n_\text{steps}$. Here, left layer is oriented as in Fig. 1(a) of the main text and right-layer is rotated in the $x-y$ plane by angle $\theta$.}
\end{figure}

\begin{figure}
\includegraphics[width=\columnwidth]{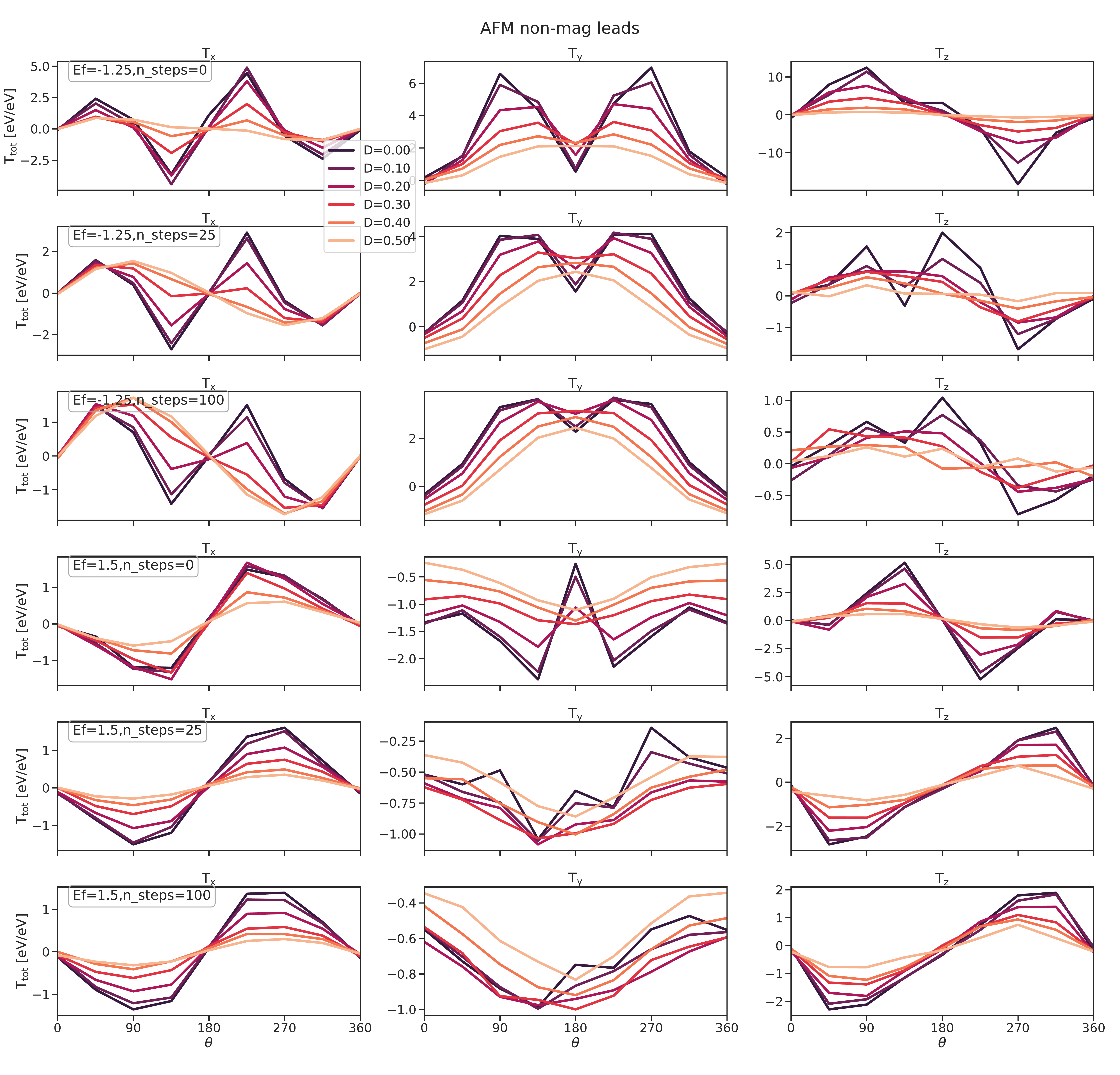}%
\caption{\label{fig:theta_dependence_NCAFM_nonmag}  The dependence of the total torque in the right magnetic region on the relative orientation of the two magnetic layers in the non-collinear AFM junction with non-magnetic leads for various values of $E_F$ and $n_\text{steps}$. Here, left layer is oriented as in Fig. 1(a) of the main text and right-layer is rotated in the $x-y$ plane by angle $\theta$.}
\end{figure}

\begin{figure}
\includegraphics[width=\columnwidth]{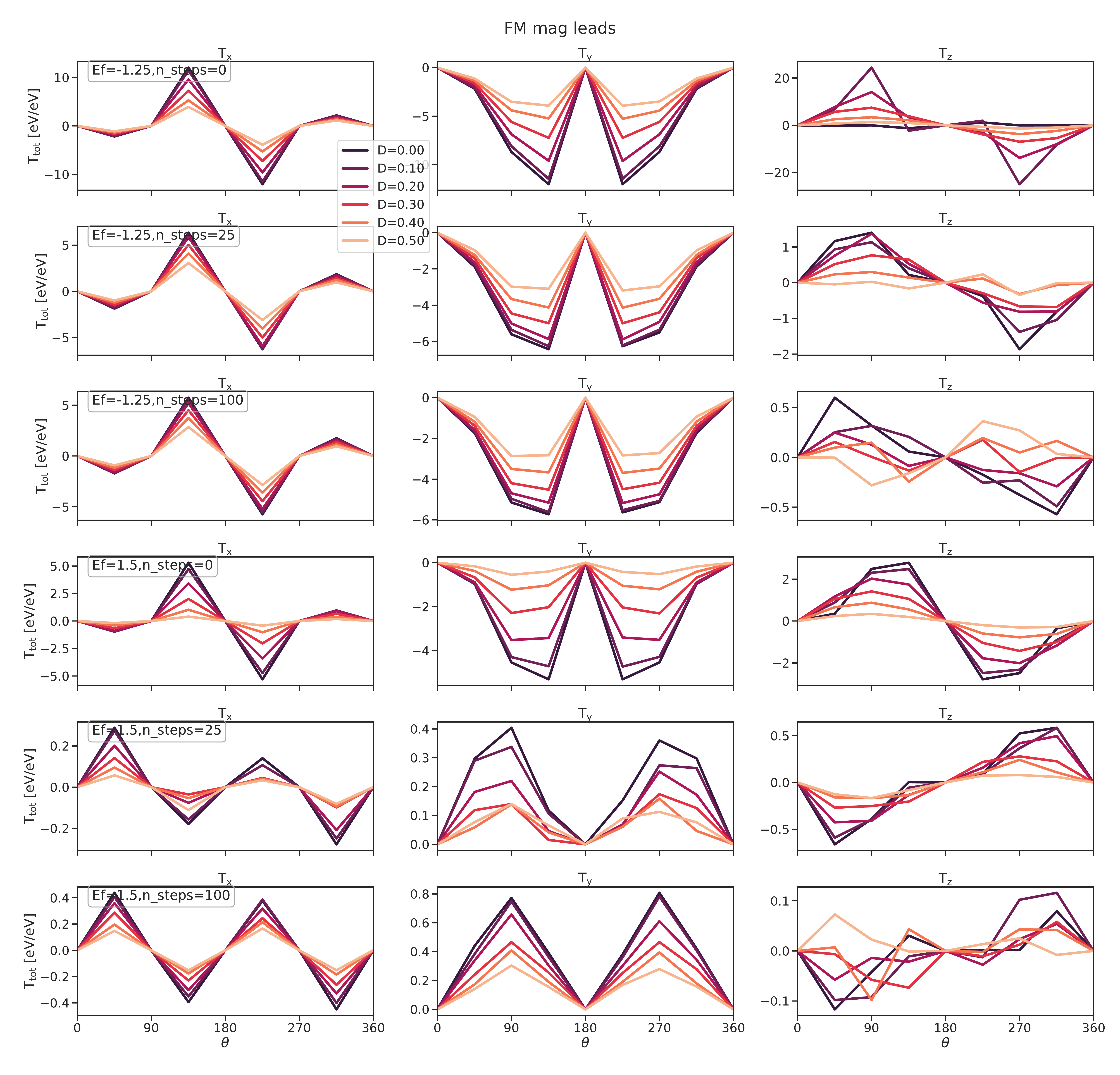}%
\caption{\label{fig:theta_dependence_FM_mag} The dependence of the total torque in the right magnetic region on the relative orientation of the two magnetic layers in the FM junction with magnetic leads for various values of $E_F$ and $n_\text{steps}$. Here, left layer is oriented along $y$ direction and right-layer is rotated in the $x-y$ plane by angle $\theta$.}
\end{figure}

\begin{figure}
\includegraphics[width=\columnwidth]{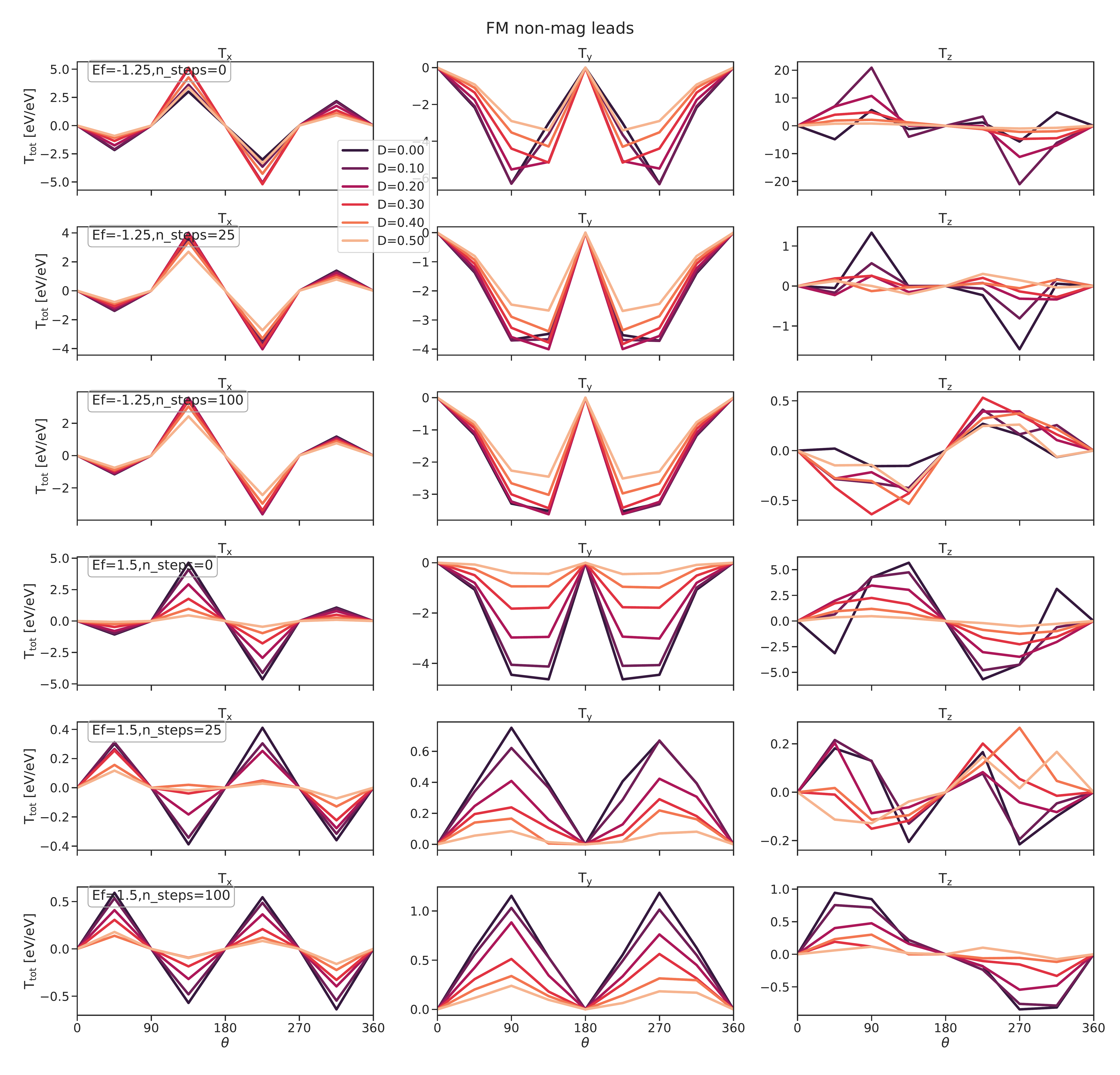}%
\caption{\label{fig:theta_dependence_FM_nonmag} The dependence of the total torque in the right magnetic region on the relative orientation of the two magnetic layers in the FM junction with non-magnetic leads for various values of $E_F$ and $n_\text{steps}$. Here, left layer is oriented along $y$ direction and right-layer is rotated in the $x-y$ plane by angle $\theta$.}
\end{figure}

In Figs. \ref{fig:theta_dependence_NCAFM_mag}-\ref{fig:theta_dependence_FM_nonmag} we give the angular dependence of the total total in the right magnetic layer for various values of $E_F$, $n_\text{steps}$, $D$ and for both magnetic and non-magnetic leads. We find that in most cases the angular dependence is not strongly sensitive to disorder.  In general we find that for the FM junction there is not much difference between the magnetic and non-magnetic leads case. In contrast, for the AFM junction, the difference between magnetic and non-magnetic leads is significant. The angular dependence in the non-magnetic leads case is more complicated, probably because the system contains twice as many interfaces.  The angular dependence is nevertheless still in most cases similar to the case of magnetic leads. The only exception is when $E_F = -1.25\ \text{eV}$ and $D \lessapprox 0.3 \text{eV}$, where we see angular dependence that resembles that of the antidamping torque in ferromagnets. 

\clearpage

\bibliography{refs}